\documentclass[journal,twocolumn]{IEEEtran}
\usepackage[dvipdfmx]{graphicx}

\usepackage{amsmath}
\usepackage{amssymb}
\usepackage{color}
\DeclareMathAlphabet{\bm}{OML}{cmm}{b}{it}
\newtheorem{theorem}{Theorem}
\newtheorem{lemma}[theorem]{Lemma}

\newtheorem{corollary}[theorem]{Corollary}
\newtheorem{remark}[theorem]{Remark}
\newtheorem{proposition}[theorem]{Proposition}
\newcommand{\qed}{\hfill \IEEEQED}

\newcommand{\markov}{\leftrightarrow}

\newcommand{\rom}[1]{\mathrm{#1}}

\allowdisplaybreaks[2]

\begin{document}

\title{The Rate-Distortion Function for Product of Two 
Sources with Side-Information at Decoders\thanks{Part of this paper was presented at 2011 IEEE International
Symposium on Information Theory, Saint Petersburg, Russia.}}

\author{Shun~Watanabe~\IEEEmembership{Member,~IEEE}       
\thanks{The author is with the Department
of Information Science and Intelligent Systems, 
University of Tokushima,
2-1, Minami-josanjima, Tokushima
770-8506, Japan, and with the Institute for System Research, University of Maryland, 
College Park, MD 20742, USA, 
e-mail:shun-wata@is.tokushima-u.ac.jp.}

\thanks{Manuscript received ; revised }}

\markboth{Journal of \LaTeX\ Class Files,~Vol.~6, No.~1, January~2007}%
{Shell \MakeLowercase{\textit{et al.}}: Bare Demo of IEEEtran.cls for Journals}

\maketitle
\begin{abstract}
This paper investigates a lossy 
source coding problem in which two decoders
can access their side-information respectively.
The correlated sources are a product of two component 
correlated sources, and we 
exclusively investigate the case such that each component
is degraded. We show the rate-distortion function
for that case, and give the following observations. 
When the components are degraded
in matched order, the rate distortion function of
the product sources is equal to the sum of the component-wise
rate distortion functions. On the other hand, 
the former is strictly smaller than the latter when the
component sources are degraded in mismatched order.
The converse proof for the mismatched case is motivated by the enhancement
technique used for broadcast channels. 
For binary Hamming and Gaussian examples, we evaluate
the rate-distortion functions. 
\end{abstract}


\IEEEpeerreviewmaketitle

\section{Introduction}

The source coding problem for correlated
sources has been regarded as an important
research area in information theory, and 
various types of coding problems were
studied so far (e.g.~\cite{slepian:73, wyner:76, berger:78, 
csiszar-korner:81, elgamal-kim-book}).
In particular, our focus in this paper is the lossy coding problem
posed by Heegard and Berger \cite{heegard:85}.

\begin{figure}[t]
\centering
\includegraphics[width=\linewidth]{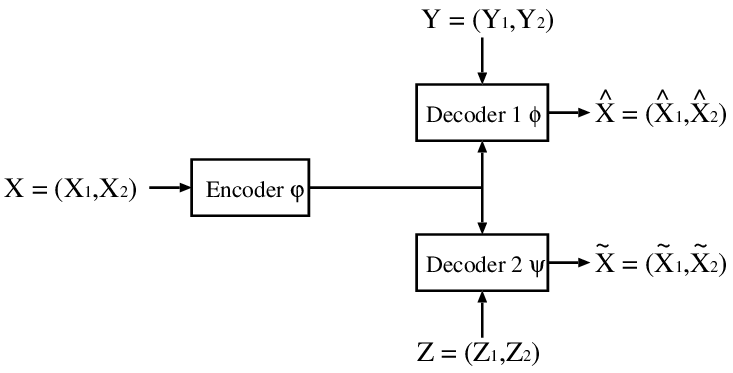}
\caption{The coding system investigated in this paper.}
\label{Fig:scenario}
\end{figure}

In the problem, there is one encoder and multiple decoders
(see Fig.~\ref{Fig:scenario}). In 
this paper, we only treat the case with two decoders.
The encoder sends an encoded version of principal source $X$.
The decoder 1 reproduces the principal source within prescribed distortion level
by the help of side-information $Y$, and 
the decoder 2 reproduces the principal source within prescribed distortion level
by the help of side-information $Z$. 

In this setting, Heegard and Berger showed an upper bound on the
rate distortion function.
They also showed that the upper bound is
tight if the side-information is degraded, i.e.,
$X$, $Y$, and $Z$ form a Markov chain in this order.
So far, there is no 
conclusive result, i.e., an upper bound and a lower bound
coincide, without the degraded assumption, and whether
Heegard and Berger's upper bound is tight or not for
non-degraded cases has been a long-standing open problem\footnote{Sgarro's problem \cite{sgarro:77} can be
regarded as a lossless special case of Heegard and Berger's problem, and his result \cite[Theorem 1]{sgarro:77} holds
without the degraded assumption.}.

In \cite{steinberg:04}, Steinberg and Merhav investigated 
the successive refinement for the Wyner-Ziv problem, which
is a generalization of Heegard and Berger's problem.
In \cite{tian:07}, Tian and Diggavi investigated the multistage
successive refinement for the Wyner-Ziv problem.
In these literatures \cite{steinberg:04, tian:07}, the side-information is
assumed to be degraded. In \cite{tian:08}, Tian and Diggavi also 
investigated the side-information scalable source coding, in which
the side-information is reversely degraded with respect to the successive refinement. 
When the refinement layer's rate of the side-information scalable source coding is $0$, 
it is nothing but Heegard and Berger's problem. In such a case, there is no difference
between the degraded and the reversely degraded.

In order to provide some insight to Heegard and Berger's problem, we 
investigate a special case of this problem
in this paper. Specifically, we consider the case such that
the correlated sources $(X,Y,Z)$ is a cartesian product of two components
correlated sources $(X_1,Y_1,Z_1)$ and $(X_2,Y_2,Z_2)$ and the
components are independent of each other
(see Fig.~\ref{Fig:scenario}). 
Furthermore, we exclusively consider the case
such that each component is degraded, i.e., either
\begin{eqnarray}
\begin{array}{l}
X_1 \markov Y_1 \markov Z_1, \\
X_2 \markov Y_2 \markov Z_2
\end{array}
\label{eq:matched-degraded}
\end{eqnarray}
or 
\begin{eqnarray}
\begin{array}{l}
X_1 \markov Y_1 \markov Z_1, \\
X_2 \markov Z_2 \markov Y_2
\end{array}
\label{eq:unmatched-degraded}
\end{eqnarray}
is satisfied, where $A \markov B \markov C$ represents that
the random variables $(A,B,C)$ form Markov chain in this order.

When (\ref{eq:matched-degraded}) is satisfied, the joint sources
$(X,Y,Z)$ are degraded. Thus, Heegard and Berger's result 
suggests that their upper bound is tight.
On the otherhand, when (\ref{eq:unmatched-degraded}) is
satisfied, the joint sources are not degraded. Thus, whether
Heegard and Berger's upper bound is tight or not is unclear so far.
In this paper, we show that the upper bound is tight whenever \eqref{eq:unmatched-degraded} holds
by finding a tight lower bound (a converse), 
i.e., we characterize the rate-distortion function.
To the best of the author's knowledge, this is the first
example such that the rate-distortion function is characterized 
without the degraded assumption\footnote{At the same time as the first version of this paper
appeared in the conference, Timo {\em et.~al.} also showed 
the rate-distortion function for some special cases of the
lossy complementary delivery problem \cite{timo:11} (see also \cite{timo:12}), which
can be also regarded as special cases of non-degraded Heegard and Berger's problem.
More specifically, Timo {\em et.~al.} solved the lossy complementary delivery problem
for the binary symmetric sources with Hamming
distortion measures, and general sources with small distortions and Hamming distortion
measures. Recently, Timo {\em et.~al.} also solved another special case
of Heegard and Berger's problem by introducing the conditional less noisy condition,
which subsumes the degraded condition \cite{timo:12b, timo:12c}.}.

\begin{figure}[t]
\centering
\includegraphics[width=\linewidth]{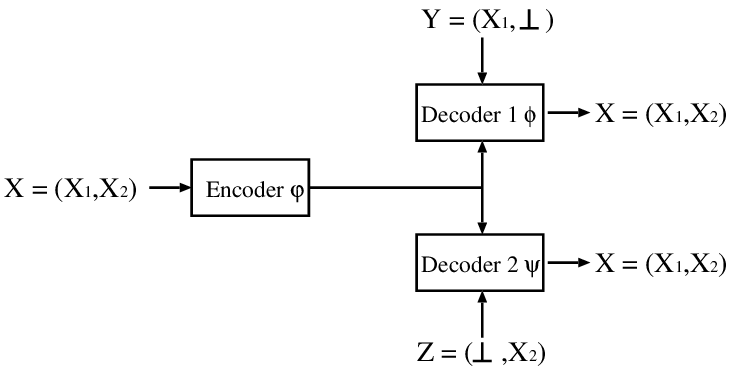}
\caption{An intuitive example such that the rate-distortion function
(with distortion $0$)
for the product source is strictly smaller than the summation of
the component-wise rate distortion functions. $X_1$ and $X_2$ are independent
uniform binary random variables, and $\perp$ represent a constant random 
 variable. When two components are encoded and decoded separately, $1$ bit
must be sent for each components, which means $2$ bits must be sent
to reproduce $(X_1,X_2)$ at both decoders. On the otherhand, if the 
 encoder sends $X_1 \oplus X_2$, then both decoders can reproduce $(X_1,X_2)$
as in the network coding \cite{ahlswede:00}.
Thus, when the components are encoded and decoded jointly, $1$ bit
suffices for the decoders to reproduce $(X_1,X_2)$. 
}
\label{Fig:example}
\end{figure}

The problem setting treated in this paper is interesting not only because 
we can obtain a conclusive result, but it is also interesting by the following reason.
Since the component correlated sources in our problem setting are independent of each other,
one might think that a combination of the component-wise optimal scheme is optimal in total
and the rate-distortion function of our problem setting is just the summation of the component-wise
rate distortion functions. However, this is not the case, i.e.,
the rate distortion function of product sources can be
strictly smaller than the summation of the component-wise
rate distortion functions even though the components
are independent of each other. To explain this fact intuitively,
let us consider an example
illustrated in Fig.~\ref{Fig:example}. When two components are encoded and decoded separately, $1$ bit
must be sent for each components, which means $2$ bits must be sent
to reproduce $(X_1,X_2)$ at both decoders. On the otherhand, if the 
 encoder sends $X_1 \oplus X_2$, then both decoders can reproduce $(X_1,X_2)$
as in the network coding \cite{ahlswede:00}\footnote{This example can be also regararded
as a special case of the complementary delivery problem. The relationship between the
complementary delivery problem and the network coding was pointed out in \cite{kimura:09}. A similar example
was also investigated in \cite[Example 1]{timo:11} (see also \cite[Example 1]{timo:12}) as an 
example of the lossy complementary delivery problem.}.
Thus, when the components are encoded and decoded jointly, $1$ bit
suffices for the decoders to reproduce $(X_1,X_2)$.
As we can find from this example, the rate distortion function of product
sources is not trivial, and it is interesting
to characterize the rate distortion function for our problem setting.

It should be noted that the present work is
motivated by the results on product of
two broadcast channels by Poltyrev \cite{poltyrev:77} and El Gamal \cite{elgamal:80}.
The broadcast channel \cite{cover} is also a long-standing open problem
in the network information theory even for two receivers. 
When there is an ordering between the two receivers (such as 
degraded, less noisy, and more capable), then
conclusive results have been obtained \cite{gallager:74, ahlswede:75, korner:75, elgamal:79, nair:10}.
Poltyrev and El~Gamal's conclusive results are few examples without such orderings.
The result in this paper can be regarded as a source coding counterpart of
Poltyrev and El~Gamal's results. However, there is a subtlety of distortions
in our problem setting that do not exist in the broadcast channel.

Recently, Weingarten {\em et.~al.}~ solved the capacity 
region of the MIMO Gaussian broadcast channel \cite{weingarten:06}.
The MIMO Gaussian broadcast channel is not degraded in general.
In \cite{weingarten:06}, the authors introduced a technique called enhancement.
There are two roles for the enhancement in the converse proof of the MIMO
Gaussian broadcast channel. One of them is a reduction of a 
MIMO non-degraded Gaussian broadcast channel to a MIMO degraded Gaussian broadcast channel.
As was pointed out in \cite[Section 9.4]{elgamal-kim-book}, Poltyrev's result \cite{poltyrev:77}
can be also derived by a straightforward application of the enhancement argument.
An application of the enhancement argument to our problem will be also discussed in this paper.
Actually, it turns out that a lower bound on the rate-distortion function derived by
a straightforward application of the enhancement argument is loose in general.

The rest of the paper is organized as follows.
In Section \ref{sec:preliminaries}, we explain the
problem setting treated in this paper, and also
explain known results.
In Section \ref{sec:main}, we show our main result
and its proof.
In Section \ref{sec:exmaple}, we show the binary Hamming example and 
the Gaussian example.

\section{Preliminaries}
\label{sec:preliminaries}

In this section,
we formally define the problem setup and review Heegard and Berger's results \cite{heegard:85}.

Let 
$(X,Y,Z) = ((X_1,X_2),(Y_1,Y_2),(Z_1,Z_2))$
be product of correlated  sources, i.e., components
$(X_1,Y_1,Z_1)$ and $(X_2,Y_2,Z_2)$ are 
independent of each other. 
The alphabet of the sources are denoted by
${\cal X} = {\cal X}_1 \times {\cal X}_2$,
${\cal Y} = {\cal Y}_1 \times {\cal Y}_2$,
and ${\cal Z} = {\cal Z}_1 \times {\cal Z}_2$
respectively, where we assume that these alphabets are finite
unless otherwise specified in the Gaussian example.
Let 
$(X^n,Y^n,Z^n)$ be $n$ independent and identically
distributed copies of $(X,Y,Z)$.

Let
$\hat{{\cal X}}_1$, $\hat{{\cal X}}_2$,
$\tilde{{\cal X}}_1$, and $\tilde{{\cal X}}_2$
be reproduction alphabets, and for $i=1,2$ let
\begin{eqnarray}
\hat{d}_i:{\cal X}_i \times \hat{{\cal X}}_i &\to& [0,\infty), 
\label{eq:distortion-measure-1} \\
\tilde{d}_i:{\cal X}_i \times \tilde{{\cal X}}_i
 &\to& [0,\infty)
  \label{eq:distortion-measure-2}
\end{eqnarray}
be distortion measures. 
Then, let
\begin{eqnarray*}
\hat{d}_{\rom{sum}}(x_1,x_2,\hat{x}_1,\hat{x}_2) &=&  \hat{d}_1(x_1,\hat{x}_1) + \hat{d}_2(x_2,\hat{x}_2), \\
\tilde{d}_{\rom{sum}}(x_1,x_2,\tilde{x}_1,\tilde{x}_2) &=& \tilde{d}_1(x_1,\tilde{x}_1) + \tilde{d}_2(x_2,\tilde{x}_2),
\end{eqnarray*}
be the sum distortion measures. 

For blocklength $n$, the coding system treated in this paper
consists of one encoder\footnote{Since it is obvious from the context, we omit subscript $n$ from the encoder, the decoders,
and the message size to simplify the notations.}
\begin{eqnarray*}
\varphi:{\cal X}^n \to \{1,\ldots,M \}
\end{eqnarray*}
and two decoders 
\begin{eqnarray*}
\phi:\{1,\ldots,M \} \times {\cal Y}^n 
\to \hat{{\cal X}}_1^n \times \hat{{\cal X}}_2^n
\end{eqnarray*}
and
\begin{eqnarray*}
\psi:\{1,\ldots,M \} \times {\cal Z}^n
\to \tilde{{\cal X}}_1^n \times \tilde{{\cal X}}_2^n.
\end{eqnarray*}
For quadruplet $\bm{D} = (\hat{D}_1,\hat{D}_2,\tilde{D}_1,\tilde{D}_2)$,
rate $R$ is said to be 
$\bm{D}$-achievable
if, for each $\gamma > 0$, there exists a code
$(\varphi,\phi,\psi)$ with a sufficiently large blocklength $n$ such that
\begin{eqnarray*}
\frac{1}{n} \log M \le R + \gamma 
\end{eqnarray*}
and 
\begin{eqnarray}
\frac{1}{n} \sum_{t=1}^n \mathbb{E}[\hat{d}_i(X_{it},\hat{X}_{it})]
 &\le& \hat{D}_i + \gamma , 
 \label{eq:distortion-constraint-1} \\
\frac{1}{n} \sum_{t=1}^n \mathbb{E}[\tilde{d}_i(X_{it},\tilde{X}_{it})]
 &\le& \tilde{D}_i + \gamma
 \label{eq:distortion-constraint-2}
\end{eqnarray}
for $i=1,2$ 
are satisfied, where 
$(\hat{X}_1^n,\hat{X}_2^n) = \phi(\varphi(X^n),Y^n)$
and 
$(\tilde{X}_2^n,\tilde{X}_2^n) = \psi(\varphi(X^n),Z^n)$.
Then, the rate-distortion function is defined as
\begin{eqnarray*}
R(\bm{D}) := \inf\{ R : \mbox{$R$ is $\bm{D}$-achievable} \}.
\end{eqnarray*}

Note that we place the individual distortion constraints in \eqref{eq:distortion-constraint-1} and \eqref{eq:distortion-constraint-2},
which are slightly different from those in the original Heegard and Beger's problem \cite{heegard:85}.
By replacing (\ref{eq:distortion-constraint-1}) and (\ref{eq:distortion-constraint-2}) with
\begin{eqnarray*}
\frac{1}{n} \sum_{t=1}^n \mathbb{E}[\hat{d}_{\rom{sum}}(X_{1t},X_{2t},\hat{X}_{1t},\hat{X}_{2t})]
 &\le& \hat{D} + \gamma ,  \\
\frac{1}{n} \sum_{t=1}^n \mathbb{E}[\tilde{d}_{\rom{sum}}(X_{1t},X_{2t},\tilde{X}_{1t},\tilde{X}_{2t})]
 &\le& \tilde{D} + \gamma
\end{eqnarray*}
respectively, we can define the rate-distortion function $R_{\rom{sum}}(\hat{D},\tilde{D})$
for the sum distortions. Since the sum distortions are special cases
of joint distortions, they are special cases of \cite{heegard:85}.

From the definitions, we obviously have
\begin{eqnarray}
\lefteqn{ R_{\rom{sum}}(\hat{D},\tilde{D}) } \nonumber \\
&\le& \min\{ R(\bm{D}): \hat{D}_1 +   \hat{D}_2 \le \hat{D}, \tilde{D}_1 +  \tilde{D}_2 \le \tilde{D} \}.
\label{eq:relation-sum-distortion}
\end{eqnarray}
When \eqref{eq:matched-degraded} or \eqref{eq:unmatched-degraded} hold,
the opposite inequality can be also proved
via the single letter characterization (see Proposition \ref{proposition-matched-case} 
and Theorem \ref{theorem-main})\footnote{It is not clear whether the opposite
inequality hold or not in general.}.

\begin{remark}
\label{remark:joint-distortion}
We can also define the rate-distortion function $R(\hat{D},\tilde{D})$ for general
joint distortions $\hat{d}$ and $\tilde{d}$. 
The single letter characterization of $R(\hat{D},\tilde{D})$ under the 
condition of (\ref{eq:matched-degraded}) can be derived from
\cite{heegard:85}. 
However, under the condition of (\ref{eq:unmatched-degraded}),
the single letter characterization of $R(\hat{D},\tilde{D})$
is not clear (see Remark \ref{remark:converse:sum-distortion}).
\end{remark}
\begin{remark}
It should be noted that the results in this paper can be easily extended to the weighted sum distortion measures
\begin{eqnarray*}
\hat{d}_{\rom{wsum}}(x_1,x_2,\hat{x}_1,\hat{x}_2) &=& \hat{\alpha} \hat{d}_1(x_1,\hat{x}_1) + \hat{\beta} \hat{d}_2(x_2,\hat{x}_2), \\
\tilde{d}_{\rom{wsum}}(x_1,x_2,\tilde{x}_1,\tilde{x}_2) &=& \tilde{\alpha} \tilde{d}_1(x_1,\tilde{x}_1) + \tilde{\beta} \tilde{d}_2(x_2,\tilde{x}_2),
\end{eqnarray*}
for some $\hat{\alpha}, \hat{\beta}, \tilde{\alpha},\tilde{\beta} \ge 0$.
\end{remark}

In \cite{heegard:85}, Heegard and Berger showed an
upper bound on the rate-distortion function.
\begin{proposition}
\label{proposition:heegard-direct}
(\cite[Theorem 2]{heegard:85}\footnote{Proposition \ref{proposition:heegard-direct} is a slight modification of
\cite[Theorem 2]{heegard:85} to component distortion functions.
The third condition, i.e., the cardinality bound was not
stated in \cite[Theorem 2]{heegard:85}, and first shown in \cite[Example 2]{timo:11b}. Since our problem in
this paper places individual distortion constraints on each component of the product source, we have to increase
the cardinalities with respect to \cite[Example 2]{timo:11b}.
})
Let $(W, \hat{U},\tilde{U})$ be auxiliary random variables satisfying
\begin{enumerate}
\item $(W, \hat{U},\tilde{U}) \markov X \markov (Y,Z)$.
\item There exist functions $\hat{X}_i^\prime(W,\hat{U},Y)$
and $\tilde{X}^\prime_i(W,\tilde{U},Z)$ such that
$\mathbb{E}[\hat{d}_i(X_i,\hat{X}_i^\prime)] \le \hat{D}_i$
and $\mathbb{E}[\tilde{d}_i(X_i,\tilde{X}_i^\prime)] \le \tilde{D}_i$
for $i = 1,2$.

\item $|{\cal W}| \le |{\cal X}| + 7$, $|\hat{{\cal U}}| \le |{\cal X}| \cdot |{\cal W}| + 2$,
and $|\tilde{{\cal U}}| \le |{\cal X}| \cdot |{\cal W}| + 2$, where ${\cal W}$, $\hat{{\cal U}}$,
and $\tilde{{\cal U}}$ are alphabets of $W$, $\hat{U}$, and $\tilde{U}$ respectively.

\end{enumerate}
Then, we have
\begin{eqnarray*}
R(\bm{D}) &\le& \max\{ I(W; X|Y), I(W; X|Z) \} \\
&&~~+ I(\hat{U}; X| Y, W) + I(\tilde{U}; X| Z, W).
\end{eqnarray*}
\end{proposition}

\begin{remark}
In \cite{heegard:85}, Heegard and Berger also showed an
upper bound on the rate-distortion function for
more than three decoders. However, Timo {\em et.~al.}~pointed out that
the statement of \cite[Theorem 2]{heegard:85} for more than three decoders is invalid,
and only the statement for two decoders is valid \cite{timo:11b}.
In \cite{timo:11b}, they also showed 
a corrected upper bound on the rate-distortion function
for more than three decoders.
\end{remark}

When the component sources are degraded 
in matched order, i.e.,
\begin{eqnarray}
\begin{array}{l}
X_1 \markov Y_1 \markov Z_1, \\
X_2 \markov Y_2 \markov Z_2
\end{array}
\label{eq:matched-degraded-redescription}
\end{eqnarray}
are satisfied, then the joint sources
$(X,Y,Z)$ are degraded, i.e.,
\begin{eqnarray*}
X \markov Y \markov Z.
\end{eqnarray*}
For the degraded sources, Heegard and Berger \cite{heegard:85}
showed that the upper bound in Proposition \ref{proposition:heegard-direct}
is tight. In particular for product of two sources, we have the following statement.
\begin{proposition}
\label{proposition-matched-case}
(\cite[Theorem 3]{heegard:85})
If the components sources are degraded
in matched order, i.e., (\ref{eq:matched-degraded-redescription})
is satisfied, then we have
\begin{eqnarray*}
R(\bm{D}) &=& R^*(\bm{D}) \\ &:=& \min[
 I(W_1;X_1|Z_1) 
 + I(U_1;X_1|Y_1,W_1) \\
&& + I(W_2;X_2|Z_2) + I(U_2;X_2|Y_2,W_2)],
\end{eqnarray*}
where the minimization is taken over 
all auxiliary random variables
$W_1,W_2,U_1,U_2$ satisfying the following:
\begin{enumerate}
\item $(W_i,U_i) \markov X_i \markov (Y_i, Z_i)$ for $i=1,2$.
\item $(W_1,U_1,X_1,Y_1,Z_1)$ and $(W_2,U_2,X_2,Y_2,Z_2)$ are independent of each other.
\item There exist functions $\hat{X}_i(W_i,U_i,Y_i)$
and $\tilde{X}_i(W_i,Z_i)$ such that 
$\mathbb{E}[\hat{d}_i(X_i,\hat{X}_i)] \le \hat{D}_i$
and $\mathbb{E}[\tilde{d}_i(X_i,\tilde{X}_i)] \le \tilde{D}_i$
for $i = 1,2$.
\item $|{\cal W}_i| \le |{\cal X}_i| + 2$ and 
$|{\cal U}_i| \le (|{\cal X}_i| + 1)^2$ for $i=1,2$, where
${\cal W}_i$ and ${\cal U}_i$ are alphabets of $W_i$ and $U_i$
respectively.
\end{enumerate}
Furthermore, we also have
\begin{eqnarray*}
\lefteqn{ R_{\rom{sum}}(\hat{D},\tilde{D}) } \\
&=& \min\{ R^*(\bm{D}): \hat{D}_1 + \hat{D}_2 \le \hat{D}, \tilde{D}_1 + \tilde{D}_2 \le \tilde{D} \}.
\end{eqnarray*}
\end{proposition}

\begin{remark}
Technically, the result in \cite[Theorem 3]{heegard:85} does not 
directly imply Proposition \ref{proposition-matched-case}, 
because Proposition \ref{proposition-matched-case} states
the stronger condition on the auxirially random variables, i.e.,
$(W_1,U_1,X_1,Y_1,Z_1)$ and $(W_2,U_2,X_2,Y_2,Z_2)$ are independent of each other. 
We give a proof of Proposition \ref{proposition-matched-case} in 
Appendix \ref{appendix:proof-of-prposition-matched} for readers' convenience.
\end{remark}

Note that $R^*(\bm{D})$ is nothing but the 
summation of the component-wise rate distortion
functions, i.e.,
\begin{eqnarray*}
R^*(\bm{D}) = R^*_1(\hat{D}_1,\tilde{D}_1) + R^*_2(\hat{D}_2,\tilde{D}_2),
\end{eqnarray*}
where 
\begin{eqnarray}
\label{component-rate-distortion-function}
R^*_i(\hat{D}_i,\tilde{D}_i) = \min[ I(W_i ; X_i|Z_i) + I(U_i ; X_i| Y_i,W_i)]
\end{eqnarray}
and the minimization in (\ref{component-rate-distortion-function})
is taken over all $(U_i,W_i)$ satisfying the
conditions 1, 3, and 4 in Proposition \ref{proposition-matched-case}.
This fact implies that the optimal scheme for the degraded product sources
is to combine the component-wise optimal scheme. 

When sources $(X,Y,Z)$ are not necessarily degraded, whether the upper 
bound in Proposition \ref{proposition:heegard-direct} is tight or not 
has been an open problem for a long time. In the next section, we will
show that the upper bound is tight if the component 
sources satisfy (\ref{eq:unmatched-degraded}).

\section{Main Results}
\label{sec:main}

\subsection{Statement of Results}

In this section, we consider the 
case in which the component sources
are degraded in mismatched order, i.e.,
\begin{eqnarray}
\begin{array}{l}
X_1 \markov Y_1 \markov Z_1, \\
X_2 \markov Z_2 \markov Y_2
\end{array}
\label{eq:unmatched-degraded-redescription}
\end{eqnarray}
are satisfied.
In this case, the joint sources 
$(X,Y,Z)$ are not degraded, and the rate-distortion
function $R(\bm{D})$ has not been clarified by any literatures.
The following is our main result, which will be proved in Section \ref{section:proof-of-main-theorem}.

\begin{theorem}
\label{theorem-main}
Suppose that $(X_1,Y_1,Z_1)$ and $(X_2,Y_2,Z_2)$ are independent of each other and
(\ref{eq:unmatched-degraded-redescription}) is satisfied.
Then, we have
\begin{eqnarray*}
\hspace{-5mm} R(\bm{D}) &=& R^\dagger(\bm{D}) \\
&:=& \min[ \max\{ I(W_1; X_1|Y_1) + I(W_2;X_2|Y_2), \\
 &&~~~~~~~~~~~~~I(W_1;X_1|Z_1) + I(W_2;X_2|Z_2) \} \\
 &&~~+ I(U_1;X_1|Y_1,W_1) + I(U_2;X_2|Z_2,W_2) ],
\end{eqnarray*}
where the minimization is taken over all auxiliary random variables
$W_1,W_2,U_1,U_2$ satisfying the following:
\begin{enumerate}
\item $(W_i,U_i) \markov X_i \markov (Y_i, Z_i)$ for $i=1,2$.
\item $(W_1,U_1,X_1,Y_1,Z_1)$ and $(W_2,U_2,X_2,Y_2,Z_2)$ are independent of each other.
\item There exist functions $\hat{X}_1(W_1,U_1,Y_1)$,
$\hat{X}_2(W_2,Y_2)$, $\tilde{X}_1(W_1,Z_1)$,
and $\tilde{X}_2(W_2,U_2,Z_2)$ such that 
\begin{eqnarray*}
\mathbb{E}[\hat{d}_i(X_i,\hat{X}_i)] \le \hat{D}_i
\end{eqnarray*}
and
\begin{eqnarray*} 
\mathbb{E}[\tilde{d}_i(X_i,\tilde{X}_i)] \le \tilde{D}_i
\end{eqnarray*}
for $i = 1,2$.
\item $|{\cal W}_i| \le |{\cal X}_i| + 3$ and 
$|{\cal U}_i| \le |{\cal X}_i| \cdot (|{\cal X}_i| + 3) + 1$ for $i=1,2$, where
${\cal W}_i$ and ${\cal U}_i$ are alphabets of $W_i$ and $U_i$
respectively.
\end{enumerate}
Furthermore, we also have
\begin{eqnarray*}
\lefteqn{ R_{\rom{sum}}(\hat{D},\tilde{D}) } \\
&=& \min\{ R^\dagger(\bm{D}): \hat{D}_1 + \hat{D}_2 \le \hat{D}, \tilde{D}_1 + \tilde{D}_2 \le \tilde{D} \}.
\end{eqnarray*}
\end{theorem}

When the distortion levels are all $0$,
we have the following corollary, which can be also derived
as a straightforward consequence of 
Sgarro's result \cite[Theorem 1]{sgarro:77}.

\begin{corollary}
When the distortion measures are the Hamming distortion measure and
$(\hat{D}_1,\hat{D}_2,\tilde{D}_1,\tilde{D}_2) = \mathbf{0} = (0,0,0,0)$,
we have
\begin{eqnarray*}
R(\mathbf{0}) &=& \max\{ H(X_1|Y_1) + H(X_2|Y_2), \\
 && ~~~~~~~H(X_1|Z_1) + H(X_2|Z_2) \} \\
 &=& \max\{ H(X_1,X_2| Y_1, Y_2), H(X_1,X_2| Z_1,Z_2) \}.
\end{eqnarray*}
\end{corollary}

\begin{remark}
It should be noted that 
\begin{eqnarray}
&& \hspace{-10mm} \max\{
H(X_1|Y_1) + H(X_2|Y_2), 
H(X_1|Z_1) + H(X_2|Z_2) \} \nonumber \\
&\le& \max\{ H(X_1|Y_1), H(X_1|Z_1)\} \nonumber \\
&& ~~~~~+ \max\{ H(X_2|Y_2), H(X_2|Z_2)\} \label{eq:loss-less-component-wise} \\
&=& H(X_1|Z_1) + H(X_2|Y_2) \label{eq:loss-less-component-wise-2} 
\end{eqnarray}
hold, and the equality in the inequality (\ref{eq:loss-less-component-wise})  does not necessarily hold in general,
where the equality in \eqref{eq:loss-less-component-wise-2} follows from \eqref{eq:unmatched-degraded-redescription}.
Note that (\ref{eq:loss-less-component-wise}) is the rate
that is needed when we apply Sgarro's coding scheme to each component. This fact implies that the combination of
the component-wise optimal scheme is not necessarily 
optimal even though the components are independent of
each other. This phenomenon also appears for lossy cases,
which will be exemplified in Section \ref{sec:exmaple}.  
\end{remark}

\subsection{Comparison to Scalable Source Coding}
\label{remark:exor-construction}

In \cite{tian:08}, Tian and Diggavi proposed a coding scheme
that is different from \cite{heegard:85}. 
Although joint encoding and decoding is required to achieve the
rate-distortion function given in Theorem \ref{theorem-main},
we can construct a code that achieve the rate-distortion function
from component-wise coding scheme of \cite{tian:08} in a similar 
manner as the example of Fig.~\ref{Fig:example}. 

When we apply the coding scheme of \cite{tian:08} to
the first component source $(X_1,Y_1,Z_1)$, the source
$X_1^n$ is quantized into the common description
$W_1^n$ and the private description $U_1^n$. 
Then, we apply the bin coding
to the common description $W_1^n$ at rate
\begin{eqnarray} \label{eq:remark-scalable-coding-1}
I(W_1;X_1|Y_1) = I(W_1;X_1) - I(W_1;Y_1),
\end{eqnarray}
where the rate $I(W_1;X_1)$ corresponds to the quantization rate and the rate
$I(W_1;Y_1)$ corresponds to the reduction of the rate by the bin coding.
Note that the equality in \eqref{eq:remark-scalable-coding-1} requires the
Markov condition $(W_1,U_1) \markov X_1 \markov (Y_1,Z_1)$.
Furthermore, we apply the bin coding to $W_1^n$ at extra rate
\begin{eqnarray*}
I(W_1;Y_1|Z_1) = I(W_1;Y_1) - I(W_1;Z_1).
\end{eqnarray*}
By using the first bin index $I_1$, the first decoder (with $Y_1^n$) can
reconstruct the common description $W_1^n$.
By using both the first bin index $I_1$ and the extra bin index $I_2$,
the second decoder (with $Z_1^n$) can reconstruct $W_1^n$.
After that the private description $U_1^n$ is transmitted to 
the first decoder at rate 
\begin{eqnarray*}
I(U_1;X_1|Y_1,W_1).
\end{eqnarray*}

Similarly, when we apply the coding scheme of \cite{tian:08} to the second
component source $(X_2,Y_2,Z_2)$, the source $X_2^n$ is quantized into
the common description $W_2^n$ and the private description $U_2^n$.
Then, we apply the bin coding to the common description $W_2^n$ at rates
\begin{eqnarray*}
I(W_2;X_2|Z_2) = I(W_2;X_2) - I(W_2;Z_2)
\end{eqnarray*}
and 
\begin{eqnarray*}
I(W_2;Z_2|Y_2) = I(W_2;Z_2) - I(W_2;Y_2)
\end{eqnarray*}
respectively so that the first decoder (with $Y_2^n$) can reconstruct
$W_2^n$ from both the first bin index $J_1$ and the second bin index $J_2$ and the second decoder (with $Z_2^n$)
can reconstruct $W_2^n$ from $J_1$. The private description $U_2^n$ is
also transmitted to the second decoder at rate
\begin{eqnarray*}
I(U_2;X_2|Z_2,W_2).
\end{eqnarray*}

By using the above two component-wise coding scheme, we can
construct a joint encoding and decoding scheme as follows.
First, the encoder sends $(I_1,J_1,I_2 \oplus J_2)$.
This requires the rate 
\begin{eqnarray*}
\lefteqn{ I(W_1;X_1|Y_1) + I(W_2;X_2|Z_2) } \\
&& + \max[
I(W_1;Y_1|Z_1), I(W_2;Z_2|Y_2)
].
\end{eqnarray*}
Note that the first (second) decoder can obtain
$J_2$ ($I_2$) by first reconstructing $W_1^n$ ($W_2^n$)
and then subtracting $I_2$ ($J_2$) from $I_2 \oplus J_2$.
The encoder also sends the private descriptions 
$U_1^n$ and $U_2^n$ at rates 
$I(U_1;X_1|Y_1,W_1)$ and $I(U_2;X_2|Z_2,W_2)$ respectively.
Consequently, the total rate coincides with the rate-distortion
function given in Theorem \ref{theorem-main}.

If we use a straightforward combination of 
the component-wise coding scheme, $I_2$ and $J_2$
will be transmitted separately instead of $I_2 \oplus J_2$, and the rate loss from the joint coding scheme is
\begin{eqnarray}
\label{eq:rate-loss}
\min[
I(W_1;Y_1|Z_1), I(W_2;Z_2|Y_2)
].
\end{eqnarray}

\subsection{Proof of Theorem \ref{theorem-main}}
\label{section:proof-of-main-theorem}

\subsubsection{Direct Part}

The direct part is a straightforward consequence of 
Proposition \ref{proposition:heegard-direct}.

For any auxiliary random variables $(W_1,W_2,U_1,U_2)$ satisfying
the conditions in Theorem \ref{theorem-main}, let 
\begin{eqnarray*}
W &=& (W_1,W_2), \\
\hat{U} &=& U_1, \\
\tilde{U} &=& U_2, \\
\hat{X}_1^\prime(W,\hat{U},Y) &=& \hat{X}_1(W_1,U_1,Y_1), \\
\hat{X}_2^\prime(W,\hat{U},Y) &=& \hat{X}_2(W_2,Y_2), \\
\tilde{X}_1^\prime(W,\tilde{U},Z) &=& \tilde{X}_1(W_1,Z_1), \\
\tilde{X}_2^\prime(W,\tilde{U},Z) &=& \tilde{X}_2(W_2,U_2,Z_2).
\end{eqnarray*}
Then, Proposition \ref{proposition:heegard-direct} implies
Theorem \ref{theorem-main}. 
The direct part for $R_{\rom{sum}}(\hat{D},\tilde{D})$ follows from
(\ref{eq:relation-sum-distortion}).
\qed

\subsubsection{Converse Part}
\label{subsec:proof-converse}

As we have mentioned in Section \ref{sec:preliminaries},
Heegard and Berger showed the converse coding
theorem for degraded case. In the course of the proof,
they essentially showed the following lemma, which can
be shown only for the degraded case.
Although our purpose is to show the converse coding
theorem for the non-degraded case,
we need the following lemma in our converse proof
of Theorem \ref{theorem-main}. A proof of Lemma \ref{lemma:degraded} is given 
in Appendix \ref{appendix:proof-of-lemma-degraded}. 

\begin{lemma}
\label{lemma:degraded}
Let 
\begin{eqnarray*}
(A,B,C) = ((A_1,A_2), (B_1,B_2), (C_1,C_2))
\end{eqnarray*}
be
product of correlated sources such that
$(A_1,B_1,C_1)$ and $(A_2,B_2,C_2)$ are independent of each other and
\begin{eqnarray}
\label{eq:condition-lemma}
A_i \markov B_i \markov C_i
\end{eqnarray}
for both $i=1$ and $i=2$. 
Let 
$(A^n,B^n,C^n)$ be $n$ independent identically distributed copies
of $(A,B,C)$. Then, for any (possibly stochastic) function
$T_n = f_n(A^n)$, we have
\begin{eqnarray*}
H(T_n) 
&\ge&
\sum_{t=1}^n \left[ 
 I(T_n, B_{1t}^-, C_{1t}^-,C_{1t}^+,C_2^n ; A_{1t}|C_{1t}) \right. \nonumber \\
 && + I(B_{1t}^+, B_2^n ; A_{1t} | B_{1t}, T_n, B_{1t}^-,C_{1t}^-,C_{1t}^+, 
 C_2^n) \nonumber \\
 && + I(T_n, B_1^n, B_{2t}^-, C_1^n, C_{2t}^-, C_{2t}^+ ; A_{2t} | C_{2t}) \nonumber \\
 && \left. 
 + I(B_{2t}^+ ; A_{2t} | B_{2t},  T_n, B_1^n, B_{2t}^-, C_1^n, C_{2t}^-, 
  C_{2t}^+ ) \right],
\end{eqnarray*}
where we use the notations 
$B_{1t}^- = (B_{11},\ldots,B_{1(t-1)})$,
$B_{1t}^+ = (B_{1(t+1)},\ldots,B_{1n})$, and etc.
\end{lemma}

We now prove the converse part for $R(\bm{D})$.
Suppose that the rate $R$ is $\bm{D}$-achievable.
Then, for any $\gamma > 0$ there exists a code
$(\varphi,\phi,\psi)$ such that
\begin{eqnarray}
\frac{1}{n} H(S_n) \le \frac{1}{n} \log M \le R + \gamma
\label{eq:proof-0}
\end{eqnarray}
and
\begin{eqnarray}
\label{eq:distortion-proof-1}
\frac{1}{n} \sum_{t=1}^n \hat{D}_{it}
 &\le& \hat{D}_i + \gamma, \\
\label{eq:distortion-proof-2}
\frac{1}{n} \sum_{t=1}^n \tilde{D}_{it}
 &\le& \tilde{D}_i + \gamma
\end{eqnarray}
for $i=1,2$ are satisfied, where we set
$S_n = \varphi(X^n)$, 
$\hat{D}_{it} = \mathbb{E}[\hat{d}_i(X_{it}, \hat{X}_{it}^n)]$
and
$\tilde{D}_{it} = \mathbb{E}[\tilde{d}_i(X_{it}, \tilde{X}_{it}^n)]$
for $(\hat{X}_1^n, \hat{X}_2^n) = \phi(\varphi(X^n),Y^n)$ 
and $(\tilde{X}_1^n, \tilde{X}_2^n)$$= \psi(\varphi(X^n), Z^n)$.

The key idea of the proof is to derive two
lower bounds on $H(S_n)$ by using
Lemma \ref{lemma:degraded} as follows.
First, let $T_n = S_n$, $(A_1,B_1,C_1) = (X_1,Y_1,Z_1)$
and $(A_2,B_2,C_2) = (X_2,Z_2,Z_2)$.
Then, since $(A,B,C)$ satisfies (\ref{eq:condition-lemma}),
we can use Lemma \ref{lemma:degraded}, and we have
\begin{eqnarray}
\lefteqn{ 
\frac{1}{n} H(S_n) } \nonumber \\
&\ge& \frac{1}{n} \sum_{t = 1}^n \left[
 I(S_n,Y_{1t}^-,Z_{1t}^-,Z_{1t}^+,Z_2^n ; X_{1t} | Z_{1t}) \right. \nonumber \\
&& + I(Y_{1t}^+ ; X_{1t} | Y_{1t}, S_n,Y_{1t}^-,Z_{1t}^-,Z_{1t}^+,Z_2^n) \nonumber \\
&& \left. + I(S_n,Y_1^n,Z_1^n,Z_{2t}^-,Z_{2t}^+ ; X_{2t} | Z_{2t}) \right] \nonumber \\
&=& \frac{1}{n} \sum_{t=1}^n \left[
 I(S_n,Y_{1t}^-,Y_2^n,Z_{1t}^-,Z_{1t}^+,Z_2^n ; X_{1t} | Z_{1t}) \right. \nonumber \\
&& + I(Y_{1t}^+ ; X_{1t} | Y_{1t}, S_n,Y_{1t}^-,Y_2^n,Z_{1t}^-,Z_{1t}^+,Z_2^n) \nonumber \\
&& \left. + I(S_n,Y_1^n,Y_{2t}^-, Y_{2t}^+, Z_1^n,Z_{2t}^-,Z_{2t}^+ ; X_{2t} | Z_{2t}) \right] 
  \label{eq:proof-1} \\
&=&  \frac{1}{n} \sum_{t=1}^n \left[
 I(W_{1t} ; X_{1t} | Z_{1t}) + I(U_{1t} ; X_{1t} | Y_{1t}, W_{1t}) \right. \nonumber \\
 && \left. + I(W_{2t}, U_{2t} ; X_{2t} | Z_{2t}) \right] \nonumber \\
&=& I(W_{1T} ; X_{1T} | Z_{1T}, T) 
 + I(U_{1T} ; X_{1T} | Y_{1T}, W_{1T}, T) \nonumber \\
 && + I(W_{2T} , U_{2T} ; X_{2T} | Z_{2T}, 
  T) \nonumber \\
&=& I(W_{1T}, T ; X_{1T} | Z_{1T})
 + I(U_{1T} ; X_{1T} | Y_{1T}, W_{1T}, T) \nonumber \\
 && + I(W_{2T } , T, U_{2T } ; X_{2T} | Z_{2T})
\label{eq:proof-2}
\end{eqnarray}
where we used the fact that 
$Y_2$ is degraded version of $Z_2$
in (\ref{eq:proof-1}), i.e., 
\begin{eqnarray*}
&& Y_2^n \markov (S_n,Y_{1t}^-,Z_1^n,Z_2^n) \markov X_{1t}, \\
&& Y_2^n \markov (S_n,Y_{1t}^-,Y_{1t},Z_{1t}^-,Z_{1t}^+,Z_2^n) \markov X_{1t}, \\
&& Y_2^n \markov (S_n,Y_1^n,Z_{1t}^-,Z_{1t}^+,Z_2^n) \markov X_{1t},
\end{eqnarray*}
and we set
\begin{eqnarray*}
W_{1t} &=& (S_n,Y_{1t}^-,Y_2^n,Z_{1t}^-,Z_{1t}^+,Z_2^n), \\
U_{1t} &=& Y_{1t}^+, \\
W_{2t} &=& (S_n,Y_1^n,Y_{2t}^-, Y_{2t}^+, Z_1^n,Z_{2t}^-), \\
U_{2t} &=& Z_{2t}^+,
\end{eqnarray*}
and $T$ is the uniform random numbers on $\{1,\ldots,n\}$
that are independent of the other random variables.
Note that $W_{1t}, U_{1t}, W_{2t}, U_{2t}$ satisfy
$(W_{it},U_{it}) \markov X_{it} \markov (Y_{it},Z_{it})$
for $i=1,2$.

Similarly, let $T_n = S_n$, $(A_1,B_1,C_1) = (X_2,Z_2,Y_2)$
and $(A_2,B_2,C_2) = (X_1,Y_1,Y_1)$. Then, since
$(A,B,C)$ satisfies (\ref{eq:condition-lemma}),
we can use Lemma \ref{lemma:degraded}, and we have
\begin{eqnarray}
\lefteqn{ 
\frac{1}{n} H(S_n) } \nonumber \\
&\ge& \frac{1}{n} \sum_{t = 1}^n \left[
I(S_n,Y_1^n,Y_{2t}^-,Y_{2t}^+,Z_{2t}^- ; X_{2t}| Y_{2t}) \right. \nonumber \\
&& + I(Z_{2t}^+ ; X_{2t} | Z_{2t}, S_n, Y_1^n,Y_{2t}^-,Y_{2t}^+,Z_{2t}^-) \nonumber \\
&& \left. + I(S_n, Y_{1t}^-,Y_{1t}^+, Y_2^n,Z_2^n ; X_{1t} | Y_{1t}) \right] \nonumber \\
&=& \frac{1}{n} \sum_{t=1}^n \left[
 I(S_n,Y_1^n,Y_{2t}^-,Y_{2t}^+,Z_1^n, Z_{2t}^- ; X_{2t}| Y_{2t}) \right. \nonumber \\
&& + I(Z_{2t}^+ ; X_{2t} | Z_{2t}, S_n, Y_1^n,Y_{2t}^-,Y_{2t}^+,Z_1^n, Z_{2t}^-) \nonumber \\
&& \left. + I(S_n, Y_{1t}^-,Y_{1t}^+, Y_2^n,Z_{1t}^-, Z_{1t}^+, Z_2^n ; X_{1t}| Y_{1t}) \right]
 \label{eq:proof-3} \\
&=& \frac{1}{n} \sum_{t=1}^n \left[
 I(W_{2t} ; X_{2t} | Y_{2t}) + I(U_{2t} ; X_{2t} | Z_{2t}, W_{2t}) \right. \nonumber \\
 && \left. + I(W_{1t}, U_{1t} ; X_{1t} | Y_{1t} ) \right], \nonumber \\
&=& I(W_{2T} ; X_{2T} | Y_{2T}, T) \nonumber 
 \\
&& + I(U_{2T} ; X_{2T} | Z_{2T}, W_{2T}, 
 T) \nonumber \\ 
&& + I(W_{1T}, U_{1T} ; X_{1T} | Y_{1T}, T) \nonumber \\
&=& I(W_{2T}, T ; X_{2T} | Y_{2T}) \nonumber 
 \\
&&  + I(U_{2T} ; X_{2T} | Z_{2T}, W_{2T}, 
 T) \nonumber \\
&& + I(W_{1T}, T, U_{1T} ; X_{1T} | Y_{1T}),
\label{eq:proof-4}
\end{eqnarray}
where we used the fact that
$Z_1$ is degraded version of $Y_1$ 
in (\ref{eq:proof-3}), i.e.,
\begin{eqnarray*}
&& Z_1^n \markov (S_n, Y_1^n, Y_2^n,Z_{2t}^-) \markov X_{2t}, \\
&& Z_1^n \markov (S_n, Y_1^n, Y_{2t}^-, Y_{2t}^+, Z_2^n) \markov X_{2t}, \\
&& Z_1^n \markov (S_n, Y_1^n, Y_{2t}^-, Y_{2t}^+, Z_{2t}^-,Z_{2t}) \markov X_{2t}.
\end{eqnarray*}

Since $(W_{1t}, U_{1t}, Y_{1t})$ and $(W_{2t}, Y_{2t})$
include $(S_n,Y_1^n, Y_2^n)$ respectively,  there exist functions
$\hat{X}_{1t}(W_{1t}, U_{1t}, Y_{1t})$ and
$\hat{X}_{2t}(W_{2t}, Y_{2t})$ satisfying
\begin{eqnarray*}
\mathbb{E}[\hat{d}_1(X_{1}, \hat{X}_{1t})] &=& \hat{D}_{1t}, \\
\mathbb{E}[\hat{d}_2(X_{2}, \hat{X}_{2t})] &=&\hat{D}_{2t}
\end{eqnarray*}
respectively. Similarly, since $(W_{1t}, Z_{1t})$ and $(W_{2t}, U_{2t}, Z_{2t})$
include $(S_n, Z_1^n,Z_2^n)$ respectively, there exist functions
$\tilde{X}_{1t}(W_{1t}, Z_{1t})$ and $\tilde{X}_{2t}(W_{2t}, U_{2t}, Z_{2t})$ satisfying 
\begin{eqnarray*}
\mathbb{E}[\tilde{d}_1(X_{1}, \tilde{X}_{1t})] &=& \tilde{D}_{1t}, \\
\mathbb{E}[\tilde{d}_2(X_{2}, \tilde{X}_{2t})] &=& \tilde{D}_{2t}
\end{eqnarray*}
respectively.
Thus, there exist functions
\begin{eqnarray*}
(\hat{X}_1(W_{1T},T,U_{1T},Y_{1T}),
\hat{X}_2(W_{2T},T,Y_{2T}))
\end{eqnarray*}
for the first decoder and functions 
\begin{eqnarray*}
(\tilde{X}_1(W_{1T}, T, Z_{1T}),
\tilde{X}_2(W_{2T}, T, U_{2T},Z_{2T}))
\end{eqnarray*}
for the second decoder satisfying
\begin{eqnarray}
\mathbb{E}[\hat{d}(X_i,\hat{X}_i)] &\le& \hat{D}_i+ \gamma
\label{eq:distortion-condition-1} \\
\mathbb{E}[\tilde{d}(X_i, \tilde{X}_i)] &\le& \tilde{D}_i+ \gamma
\label{eq:distortion-condition-2}
\end{eqnarray}
for $i=1,2$.
Thus, by combining (\ref{eq:proof-0}), (\ref{eq:proof-2}), and (\ref{eq:proof-4}),
and by taking $W_1 = (W_{1T},T)$, $U_1 = U_{1T}$,
$W_2 = (W_{2T},T)$, and $U_2 = U_{2T}$, 
we have that there exist $W_1,W_2,U_1,U_2$ satisfying 
(\ref{eq:distortion-condition-1}) 
and (\ref{eq:distortion-condition-2}) and
\begin{eqnarray*}
R(\bm{D}) 
&\ge& I(W_1; X_1|Z_1) 
+ I(U_1; X_1| Y_1, W_1) \\
&& + I(W_2, U_2; X_2|Z_2) - \gamma, \\
R(\bm{D})
&\ge& I(W_1, U_1; X_1|Y_1) \\
&& + I(W_2; X_2|Y_2) 
 + I(U_2 ; X_2|Z_2, W_2) - \gamma.
\end{eqnarray*}
Although the auxirially random variables 
$(W_1,U_1,X_1,Y_1,Z_1)$ and $(W_2,U_2,X_2,Y_2,Z_2)$ chosen above are
not necessarily independent of each other, they do not
appear together in any one term. Thus we can take
$(W_1,U_1,X_1,Y_1,Z_1)$ and $(W_2,U_2,X_2,Y_2,Z_2)$ to be independent of each other. 

By applying the cardinality bound on the auxiliary random variables, which will be proved in
Appendix \ref{appendix:cardinality-bounds}, we have
\begin{eqnarray*}
R(\bm{D}) \ge R^\dagger(\bm{D} + \gamma \mathbf{1}) - \gamma,
\end{eqnarray*}
where $\mathbf{1} = (1,1,1,1)$.
Since $\gamma > 0$ is arbitrary,
by the continuity of $R^\dagger(\bm{D})$ with respectto $\bm{D}$, we have the converse part for 
$R(\bm{D})$\footnote{Since the cardinalities of the auxiliary random
variables are bounded, $R^\dagger(\bm{D})$ can be described as a finite dimensional optimization problem and the continuity 
of $R^\dagger(\bm{D})$ with respectto $\bm{D}$ follows from the continuity of the mutual information with respect to the test channel.}
.
The converse part for $R_{\rom{sum}}(\hat{D},\tilde{D})$ can 
be proved almost in the same manner.
\qed

\begin{remark}
\label{remark:converse:sum-distortion}
In the above converse proof, we derived the independence between 
$(W_1,U_1,X_1,Y_1,Z_1)$ and $(W_2,U_2,X_2,Y_2,Z_2)$ by using the fact that they
do not appear together in any term one term. Thus, we 
cannot derive the independence between 
them if we employ general joint distortion measures $\hat{d}$ and $\tilde{d}$.
Without this independence, we cannot prove the matching direct part 
from Proposition \ref{proposition:heegard-direct}
because
\begin{eqnarray*}
\lefteqn{ I(W_1,W_2; X_1,X_2|Y_1,Y_2) } \\
&=& I(W_1;X_1|Y_1) + I(W_2;X_2|Y_2), \\
\lefteqn{ I(W_1,W_2; X_1,X_2|Z_1,Z_2) } \\
&=& I(W_1; X_1|Z_1) + I(W_2;X_2|Z_2)
\end{eqnarray*}
do not hold in general.
For this reason, the single letter characterization of $R(\hat{D},\tilde{D})$
is not clear.
\end{remark}


\begin{remark}
\label{remark:enhancement}
In the above converse argument, we reduced the proof to the degraded case
by setting $(A_1,B_1,C_1) = (X_1,Y_1,Z_1)$
and $(A_2,B_2,C_2) = (X_2,Z_2,Z_2)$, or by setting
$(A_1,B_1,C_1) = (X_2,Z_2,Y_2)$
and $(A_2,B_2,C_2) = (X_1,Y_1,Y_1)$.
This reduction argument is motivated by the enhancement technique introduced by 
Weingarten {\em et. al.} \cite{weingarten:06}, in which the converse proof of
the MIMO (not necessarily degraded) broadcast channel was reduced to
that of the MIMO degraded broadcast channel. This kind of argument was implicitly 
used in \cite{elgamal:80}. As is pointed out in \cite[Section 9.4]{elgamal-kim-book},
the result in \cite{poltyrev:77} can be obtained by a straightforward application
of the enhancement argument.

It should be noted that the following straightforward application
of the enhancement argument gives only loose converse in our problem.
Suppose that $((X_1,X_2),(Y_1,Y_2),(Z_1,Z_2))$ satisfies the Markov 
conditions in (\ref{eq:unmatched-degraded-redescription}), and let
\begin{eqnarray*}
R(\bm{D}|((X_1,X_2),(Y_1,Y_2),(Z_1,Z_2)))
\end{eqnarray*}
 be the rate-distortion function
for this source. Let 
\begin{eqnarray*}
R(\bm{D}|((X_1,X_2),(Y_1,Z_2),(Z_1,Z_2)))
\end{eqnarray*}
and 
\begin{eqnarray*}
R(\bm{D}|((X_1,X_2),(Y_1,Y_2),(Y_1,Z_2)))
\end{eqnarray*}
be the rate-distortion
functions for the enhanced sources respectively.
Then,  we have
\begin{eqnarray}
\lefteqn{ R(\bm{D}|((X_1,X_2),(Y_1,Y_2),(Z_1,Z_2))) } \nonumber \\
&\ge& \max\{ R(\bm{D}|((X_1,X_2),(Y_1,Z_2),(Z_1,Z_2))), \nonumber \\
&&~~~~~~~ R(\bm{D}|((X_1,X_2),(Y_1,Y_2),(Y_1,Z_2))) \}.
\label{eq:loose-enhanced-bound}
\end{eqnarray}
As will be exemplified in Section \ref{sec:gaussian},
this lower bound is loose in general.
\end{remark}

\section{Examples}
\label{sec:exmaple}

To illustrate our main result, we consider 
a binary example and a Gaussian example.

\subsection{Binary Example}
\label{sec:binary}

In this section, we evaluate the 
rate distortion function for the binary Hamming example.
We first review some known result of the binary Hamming version of
the rate-distortion function where the side-information may be absent \cite{heegard:85}. 
This result will be used to
investigate the rate-distortion function for product of two binary sources.

Let $X$ be the uniform binary source, and let $Y$ be the output
of the binary symmetric channel with crossover probability $p < \frac{1}{2}$, where the
input is $X$. Let $Z$ be a constant, and let $d$ be the Hamming distortion measure.
The rate-distortion function of this situation is given by 
\begin{eqnarray}
\label{eq:rd-may-be-absent}
R_{b}(\hat{D},\tilde{D}) = \min[ I(W;X) + I(U;X|W,Y) ],
\end{eqnarray}
where the minimization is taken over all auxiliary random variables $W$ and $U$
satisfying the following:
\begin{enumerate}
\item $(W,U) \markov X \markov Y$. \label{condition-single-binary-1}
\item There exist functions $\hat{X}(W,U,Y)$ and $\tilde{X}(W)$ such that 
$\mathbb{E}[d(X,\hat{X})] \le \hat{D}$ and $\mathbb{E}[d(X,\tilde{X})] \le \tilde{D}$. \label{condition-single-binary-2}
\item $|{\cal W}| \le |{\cal X}| + 2$ and $|{\cal U}| \le (|{\cal X}| + 2)^2$.
\end{enumerate}

An explicit form of $R_b(\hat{D},\tilde{D})$ 
was first studied in \cite{heegard:85}, and a loose upper bound was obtained. 
After that, Kerpez \cite{kerpez:87}
and Fleming and Effros \cite{fleming:06} also studied this problem. 
Finally, Tian and Diggavi \cite{tian:07} derived an explicit form of $R_b(\hat{D},\tilde{D})$.

For $0 \le q \le 1$, let
\begin{eqnarray*}
G_p(q) = h(p * q) - h(q),
\end{eqnarray*}
where $h(\cdot)$ is the binary entropy function and $p * q = p (1-q) + (1-p) q$ is the binary convolution.
It was shown in \cite{kerpez:87} that the rate distortion region can be partitioned into four
subregions, three of which are degenerate.
\begin{itemize}
\item Region I: $0 \le \tilde{D} < \frac{1}{2}$ and $0 \le \hat{D} < \min\{ \tilde{D}, p\}$. In this region, $R_b(\hat{D},\tilde{D})$
is a function of both $\hat{D}$ and $\tilde{D}$, and it is the only non-degenerate case.

\item Region II: $\tilde{D} \ge \frac{1}{2}$ and $0 \le \hat{D} < p$. In this region, the common description $W$
is not needed, and $R_b(\hat{D},\tilde{D})$ reduces to the Wyner-Ziv rate-distortion function, i.e.,
\begin{eqnarray*}
R_b(\hat{D},\tilde{D}) &=& R_p^{WZ}(\hat{D}) \\
&=& \min_{(\beta,\theta):0 \le \theta \le 1, 0 \le \beta \le p, \theta \beta + (1-\theta) p = \hat{D}} [\theta G(\beta)].
\end{eqnarray*}

\item Region III: $0 \le \tilde{D} < \frac{1}{2}$ and $\hat{D} \ge \min\{ \tilde{D},p\}$. In this region, the refinement description $U$
is not needed, and $R_b(\hat{D},\tilde{D})$ reduces to the ordinary rate-distortion function, i.e.,
\begin{eqnarray*}
R_b(\hat{D},\tilde{D}) = 1 - h(\tilde{D}).
\end{eqnarray*}

\item Region IV: $\tilde{D} \ge \frac{1}{2}$ and $\hat{D} \ge p$. In this region, clearly both descriptions $W$ and $U$ can
be constant, and $R_b(\hat{D},\tilde{D}) = 0$.
\end{itemize}

\begin{figure}[t]
\centering
\includegraphics[width=0.4\linewidth]{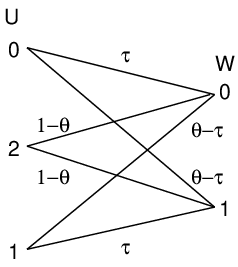}
\caption{The test channel between $U$ and $W$.}
\label{Fig:test-channel-w12}
\end{figure}

\begin{figure}[t]
\centering
\includegraphics[width=0.5\linewidth]{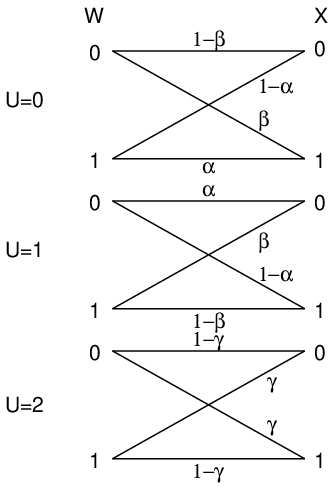}
\caption{The test channel between $W$ and $X$.}
\label{Fig:test-channel-wx}
\end{figure}

To describe the rate-distortion function for region I, we need 
to introduce some notations. 
For parameters $(D, \alpha,\beta,\theta,\tau)$ satisfying
\begin{eqnarray*}
0 \le D \le \frac{1}{2},~0 \le \alpha,\beta \le p,~ 0 \le \tau \le \theta \le 1,
\end{eqnarray*}
we define
\begin{eqnarray*}
B_p(D,\alpha, \beta, \theta,\tau) &=& (\theta - \tau) G_p(\alpha) + \tau G_p(\beta) \\
&& + (1-\theta) G_p(\gamma(D,\alpha,\beta,\theta,\tau)),
\end{eqnarray*}
where 
\begin{eqnarray*}
\gamma(D,\alpha,\beta,\theta,\tau) = \left\{
\begin{array}{ll}
\frac{D - (\theta - \tau)(1-\alpha) - \tau \beta}{1 - \theta} & \theta \neq 1 \\
\frac{1}{2} & \theta = 1
\end{array}
\right..
\end{eqnarray*}
We also define 
\begin{eqnarray*}
\lefteqn{ {\cal Q}_p(\hat{D},\tilde{D}) } \\
&=& \{ (\check{D},\alpha,\beta,\theta,\tau) : \\
&& \hspace{-4mm} (1 - \theta) p \le \check{D} - (\theta - \tau)(1-\alpha) - \tau \beta \le (1-\theta) (1-p), \\
&& \hspace{-4mm} 0 \le \tau \le \theta \le 1, \\
&& \hspace{-4mm} 0 \le \alpha, \beta \le p,\\
&& \hspace{-4mm} (\theta - \tau)\alpha + \tau \beta + (1-\theta)p \le \hat{D}, \\
&& \hspace{-4mm} \check{D} \le \tilde{D} \}.
\end{eqnarray*}
For region I, Tian and Diggavi \cite{tian:07} showed\footnote{Tian and Diggavi also showed that
the restriction to the equalities $\check{D} = \tilde{D}$ and $(\theta - \tau)\alpha + \tau \beta + (1-\theta)p = \hat{D}$
in the definition of ${\cal Q}_p(\hat{D},\tilde{D})$ does not increase the rate-distortion function. However, in the case of
the product of two sources, it is not clear whether such a restriction does not increase the rate-distortion function.}
\begin{eqnarray}
\lefteqn{ R_b(\hat{D},\tilde{D})   } \nonumber \\
&=& \hspace{-3mm} \min_{(\check{D}, \alpha,\beta,\theta,\tau) \in {\cal Q}_p(\hat{D},\tilde{D})} [ 1 - h(\check{D} * p)+    B_p(\check{D},\alpha,\beta,\theta,\tau)].
\nonumber \\
\label{eq:tian-diggavi}
\end{eqnarray}
The righthand side of (\ref{eq:rd-may-be-absent}) can be rewritten as
\begin{eqnarray}
\min[ I(W;Y) + I(U,W;X|Y)  ],
\label{eq:another-form-of-hb}
\end{eqnarray}
and $R_b(\hat{D},\tilde{D})$ is achieved by  reverse test channels 
described in Figs.~\ref{Fig:test-channel-w12} and \ref{Fig:test-channel-wx}.
Note that 
\begin{eqnarray}
&& \hspace{-15mm} \min_{(\check{D},\alpha,\beta,\theta,\tau) \in {\cal Q}_p(\hat{D},\tilde{D})} [ 
1 - h(\check{D} * p) + B_p(\check{D},\alpha,\beta,\theta,\tau) ] 
\nonumber \\
&=& R_p^{WZ}(\hat{D})
\label{eq:end-point-1}
\end{eqnarray}
for $\tilde{D} = \frac{1}{2}$ and
\begin{eqnarray}
&& \hspace{-15mm} \min_{(\check{D},\alpha,\beta,\theta,\tau) \in {\cal Q}_p(\hat{D},\tilde{D})} [
 1 - h(\check{D} * p) + B_p(\check{D},\alpha,\beta,\theta,\tau) ] \nonumber \\
&=& 1 - h(\tilde{D} * p) + G_p(\tilde{D}) \label{eq:end-point-2} \\
&=& 1 - h(\tilde{D}) \nonumber 
\end{eqnarray}
for $\hat{D} = \min\{ \tilde{D},p\}$ and $\tilde{D} \le \frac{1}{2}$, which will be 
proved in Appendix \ref{appendix-proof-end-point}.
Thus, we can also write
\begin{eqnarray*}
\lefteqn{ R_b(\hat{D},\tilde{D}) } \\
 &=&  \min_{(\check{D},\alpha,\beta,\theta,\tau) \in {\cal Q}_p(\hat{D},\tilde{D})} [ 1 - h(\check{D} * p) 
+    B_p(\check{D},\alpha,\beta,\theta,\tau)] \\
\label{eq:tian-diggavi-2}
\end{eqnarray*}
for any $(\hat{D},\tilde{D})$.


Now, we consider the rate-distortion function for product of two binary sources.
Let $X_1$ and $X_2$ be the independent uniform binary sources. 
Let $Y_1$ be the output of the binary symmetric channel with crossover probability $p_1 < \frac{1}{2}$,
where the input is $X_1$.
Let $Z_2$ be the outputs of the binary symmetric channel with crossover probability $p_2 < \frac{1}{2}$,
where the input is $X_2$. Then, let $Y_2$ and $Z_1$ be constant.  Obviously, this pair of
correlated sources satisfy the Markov conditions in (\ref{eq:unmatched-degraded-redescription}).
In this case, we have the following.
\begin{theorem}
\label{theorem:binary-hamming}
For any $\bm{D} = (\hat{D}_1,\tilde{D}_1,\hat{D}_2,\tilde{D}_2)$, we have
\begin{eqnarray}
R(\bm{D}) 
&=& \min[ \max\{ 1 - h(\check{D}_1 * p_1), 1 - h(\check{D}_2 * p_2)\} \nonumber \\
&& + B_{p_1}(\check{D}_1, \alpha_1,\beta_1,\theta_1,\tau_1) \nonumber \\
&&  + B_{p_2}(\check{D}_2,\alpha_2,\beta_2,\theta_2,\tau_2)],
\label{eq:binary-example-expocit-form}
\end{eqnarray}
where the minimizations are taken over 
\begin{eqnarray*}
(\check{D}_1,\alpha_1,\beta_1,\theta_1,\tau_1) \in {\cal Q}_{p_1}(\hat{D}_1,\tilde{D}_1)
\end{eqnarray*}
and 
\begin{eqnarray*}
(\check{D}_2, \alpha_2,\beta_2,\theta_2,\tau_2) \in {\cal Q}_{p_2}(\tilde{D}_2,\hat{D}_2)
\end{eqnarray*}
respectively.
\end{theorem}
\begin{proof}
See Appendix \ref{appendix:proof-of-theorem:binary-hamming}.
\end{proof}

In the following, for a symmetric case, we compare the rate-distortion 
function, the upper bound derived by the component-wise scheme, and the
lower bound derived by the straightforward enhancement.
Let $p_1 = p_2 = p < \frac{1}{2}$.
Let $d_c( p)$ the critical distortion \cite{wyner:76}, i.e., the distortion satisfying
\begin{eqnarray*}
\frac{G_p(d_c(p ))}{d_c( p) - p} = G^\prime_p(d_c( p)).
\end{eqnarray*}  
Let $\tilde{D}_1 = \hat{D}_2 = \bar{D}$, where 
 $d_c( p) < \bar{D} < \frac{1}{2}$.
Let $\hat{D}_1 = \tilde{D}_2 = D$.
From (\ref{eq:tian-diggavi-2}), it is clear that the summation
of the component-wise rate-distortion functions is
\begin{eqnarray*}
2 \min_{(\check{D},\alpha,\beta,\theta,\tau) \in {\cal Q}_p(D,\bar{D})} [ 1 -  h(\check{D} * p) +    B_p(\check{D},\alpha,\beta,\theta,\tau)],
\end{eqnarray*}
which is strictly larger than the joint rate-distortion function
obtained from Theorem \ref{theorem:binary-hamming}.

Suppose that $D \ge \bar{D}$.
In this case, in a similar manner as (\ref{eq:end-point-2}), 
we can show that the joint rate-distortion function is 
\begin{eqnarray*}
R(\bm{D}) = 1 - h(\bar{D} * p) + 2 G_p(\bar{D}).
\end{eqnarray*}
On the other hand, from Proposition \ref{proposition-matched-case}, 
the rate-distortion function of the source satisfying (\ref{eq:matched-degraded-redescription}) is 
the summation of the component-wise rate-distortion functions. Thus,
the lower bound in (\ref{eq:loose-enhanced-bound}) is given by
\begin{eqnarray*}
1 - h(\bar{D}*p) + G_p(\bar{D}) + R_p^{WZ}(\bar{D}).
\end{eqnarray*}
Since $R_p^{WZ}(\bar{D}) < G_p(\bar{D})$ for $d_c( p) < \bar{D} < \frac{1}{2}$,
the lower bound in (\ref{eq:loose-enhanced-bound}) is loose.

Suppose that 
$D \le  d_c( p)$.
In this case, in a similar manner as \cite[Corollary 2]{tian:07}, 
we can show that the joint rate-distortion function is 
\begin{eqnarray}
\label{eq:lower-bound-tight-case}
1 - h(\bar{D}*p) + 2 G_p(D).
\end{eqnarray}
The lower bound in (\ref{eq:loose-enhanced-bound}) coincide with 
(\ref{eq:lower-bound-tight-case}) in this case, and thus tight.

\subsection{Gaussian Example}
\label{sec:gaussian}

In this section, we evaluate the 
rate distortion function for the Gaussian example.
We consider jointly Gaussian sources $(X_i,Y_i,Z_i)$
given by
$Y_i = X_i + N_{i,y}$ and
$Z_i = X_i + N_{i,z}$.
where $N_{i,y}$ and $N_{i,z}$ are Gaussian
noises with variances $\Sigma_{i,N_y}$ and 
$\Sigma_{i,N_z}$ such that
$\Sigma_{1,N_y} < \Sigma_{1,N_z}$ and
$\Sigma_{2,N_z} < \Sigma_{2,N_y}$ respectively.
The conditional variance of $X_i$ given 
$Y_i$ is denoted by 
$\Sigma_{i,x|y}$ etc..
To avoid tedious degenerate cases, we assume that $\hat{D}_i < \Sigma_{i,x|y}$
and $\tilde{D}_i < \Sigma_{i,x|z}$ for $i=1,2$. 

In the above setting, the rate-distortion function
is given by the following theorem. The theorem can be
proved by first showing that Gaussian auxiliary random variables
suffice, and then by elementary calculation.
\begin{theorem}
We have
\begin{eqnarray*}
\lefteqn{R(\bm{D}) } \\
&=& 
\max\left[
\frac{1}{2}\log \frac{\Sigma_{1,x|y}}{ (\tilde{D}_1^{-1} - \Sigma_{1,N_z}^{-1} + \Sigma_{1,N_y}^{-1})^{-1} }
 \right.  \\
&& ~~~~~~ + \frac{1}{2}\log \frac{\Sigma_{2,x|y}}{ \hat{D}_2}, \\
&& \left. \frac{1}{2}\log \frac{\Sigma_{1,x|z}}{ \tilde{D}_1}
 + \frac{1}{2}\log \frac{\Sigma_{2,x|z}}{(\hat{D}_2^{-1} - \Sigma_{2,N_y}^{-1} + \Sigma_{2,N_z}^{-1})^{-1}}
\right] \\
&& + \frac{1}{2}\log \frac{(\tilde{D}_1^{-1} - \Sigma_{1,N_z}^{-1} + \Sigma_{1,N_y}^{-1})^{-1}}{(B_1^* + \Sigma_{1,N_y}^{-1})^{-1}} \\
&& + \frac{1}{2}\log \frac{(\hat{D}_2^{-1} - \Sigma_{2,N_y}^{-1} + \Sigma_{2,N_z}^{-1})^{-1}}{(B_2^* + \Sigma_{2,N_z}^{-1})^{-1}},
\end{eqnarray*}
where 
\begin{eqnarray*}
B_1^* &=& \max[ \tilde{D}_1^{-1} - \Sigma_{1,N_z}^{-1}, \hat{D}_1^{-1} - \Sigma_{1,N_y}^{-1}],\\
B_2^* &=& \max[ \hat{D}_2^{-1} - \Sigma_{2,N_y}^{-1}, \tilde{D}_2^{-1} - \Sigma_{2,N_z}^{-1}].
\end{eqnarray*}
\end{theorem}

Note that the component-wise rate-distortion functions are given
by 
\begin{eqnarray*}
R^*_1(\hat{D}_1,\tilde{D}_1) &=&
 \frac{1}{2} \log \frac{\Sigma_{1,x|z}}{ \tilde{D}_1} \\
&& + \frac{1}{2} \log \frac{(\tilde{D}_1^{-1} - \Sigma_{1,N_z}^{-1} + 
 \Sigma_{1,N_y}^{-1})^{-1}}{(B_1^* + 
 \Sigma_{1,N_y}^{-1})^{-1}}, \\
R^*_2(\hat{D}_2,\tilde{D}_2) &=&
 \frac{1}{2} \log \frac{\Sigma_{2,x|y}}{ \hat{D}_2} \\
&& + \frac{1}{2} \log \frac{(\hat{D}_2^{-1} - \Sigma_{2,N_y}^{-1} + \Sigma_{2,N_z}^{-1})^{-1}}{(B_2^* + 
 \Sigma_{2,N_z}^{-1})^{-1}}.
\end{eqnarray*}
By noting $\Sigma_{1,N_y} < \Sigma_{1,N_z}$
and $\Sigma_{2,N_y} > \Sigma_{2,N_z}$, we have
\begin{eqnarray*}
\frac{\tilde{D}_1^{-1} - \Sigma_{1,N_z}^{-1} + \Sigma_{1,N_y}^{-1}}{\Sigma_{1,x|y}^{-1}}
&=& \frac{\tilde{D}_1^{-1} - \Sigma_{1,N_z}^{-1} + \Sigma_{1,N_y}^{-1}}{\Sigma_{1,x|z}^{-1} - \Sigma_{1,N_z}^{-1} + \Sigma_{1,N_y}^{-1}} \\
&<& \frac{\tilde{D}_1^{-1}}{\Sigma_{1,x|z}^{-1}}
\end{eqnarray*}
and 
\begin{eqnarray*}
\frac{\hat{D}_2^{-1} - \Sigma_{2,N_y}^{-1} + \Sigma_{2,N_z}^{-1}}{\Sigma_{2,x|z}^{-1}}
&=& \frac{\hat{D}_2^{-1} - \Sigma_{2,N_y}^{-1} + \Sigma_{2,N_z}^{-1}}{\Sigma_{2,x|y}^{-1} - \Sigma_{2,N_y}^{-1} + \Sigma_{2,N_z}^{-1}} \\
&<& \frac{\hat{D}_2^{-1}}{\Sigma_{2,x|y}^{-1}}.
\end{eqnarray*}
Thus, we have
\begin{eqnarray*}
R(\bm{D}) < R_1^*(\hat{D}_1,\tilde{D}_1)
 + R_2^*(\hat{D}_2,\tilde{D}_2),
\end{eqnarray*}
which implies that the combination of the 
component-wise optimal scheme is suboptimal for
Gaussian product sources.

Next, we consider
the lower bound in (\ref{eq:loose-enhanced-bound}).
Let
\begin{eqnarray*}
D_{1,\min} &=& \min\{ \hat{D}_1, \tilde{D}_1\}, \\
D_{2,\min} &=& \min\{ \hat{D}_2, \tilde{D}_2\}.
\end{eqnarray*}
Then, using the same notations as in Remark \ref{remark:enhancement},  we have
\begin{eqnarray*}
\lefteqn{ R(\bm{D}|((X_1,X_2),(Y_1,Z_2), (Z_1,Z_2))) } \\
&=& \frac{1}{2}\log \frac{\Sigma_{1,x|z}}{\tilde{D}_1} 
  + \frac{1}{2}\log \frac{(\tilde{D}_1^{-1} - \Sigma_{1,N_Z}^{-1} + \Sigma_{1,N_y}^{-1})^{-1}}{(B_1^* + \Sigma_{1,N_y}^{-1})^{-1}} \\
  && + \frac{1}{2}\log \frac{\Sigma_{2,x|z}}{D_{2,\min}}
\end{eqnarray*}
and 
\begin{eqnarray*}
\lefteqn{ R(\bm{D}|((X_1,X_2),(Y_1,Y_2),(Y_1,Z_2)))  } \\
&=& \frac{1}{2} \log \frac{\Sigma_{1,x|y}}{D_{1,\min}} \\
&& + \frac{1}{2} \log \frac{\Sigma_{2,x|y}}{\hat{D}_2} 
 + \frac{1}{2} \log \frac{(\hat{D}_2^{-1} - \Sigma_{2,N_y}^{-1} + \Sigma_{2,N_z}^{-1})^{-1}}{(B_2^* + \Sigma_{2,N_z}^{-1})^{-1}}. 
\end{eqnarray*}
Thus, if 
\begin{eqnarray}
\label{eq:gaussian-matching-condition-enhancement-1}
\hat{D}_1 &\le& (\tilde{D}_1^{-1} - \Sigma_{1,N_z}^{-1} + \Sigma_{1,N_y}^{-1})^{-1}, \\
\tilde{D}_2 &\le& (\hat{D}_2^{-1} - \Sigma_{2,N_y}^{-1} + \Sigma_{2,N_z}^{-1})^{-1},
\label{eq:gaussian-matching-condition-enhancement-2}
\end{eqnarray}
then we have
\begin{eqnarray*}
B_1^* + \Sigma_{1,N_y}^{-1} &=& D_{1,\min}^{-1} = \hat{D}_1^{-1}, \\
B_2^* + \Sigma_{2,N_z}^{-1} &=& D_{2,\min}^{-1} = \tilde{D}_2^{-1},
\end{eqnarray*}
and the lower bound in (\ref{eq:loose-enhanced-bound}) is tight.
However, if (\ref{eq:gaussian-matching-condition-enhancement-1}) 
or (\ref{eq:gaussian-matching-condition-enhancement-2}) are not satisfied,
then the lower bound in (\ref{eq:loose-enhanced-bound}) is not necessarily tight. 

In the following, for a symmetric case, we compare the rate-distortion 
function, the upper bound derived by the component-wise scheme, and the
lower bound derived by the straightforward enhancement, i.e.,
the lower bound in (\ref{eq:loose-enhanced-bound}).
We set $\Sigma_{X_1} = \Sigma_{X_2} = \Sigma_X$,
$\Sigma_{1,N_y} = \Sigma_{2,N_z} = \Sigma_N$,
$\Sigma_{1,N_z} = \Sigma_{2,N_y} = \Sigma_{\bar{N}}$,
$\hat{D}_1 = \tilde{D}_2 = D$, and $\tilde{D}_1 = \hat{D}_2 = \bar{D}$,
where $\Sigma_N < \Sigma_{\bar{N}}$. In this case, we have
\begin{eqnarray*}
\Sigma_{1,x|y} &=& \Sigma_{2,x|z} =: \Sigma_{x|s}, \\
\Sigma_{1,x|z} &=& \Sigma_{2,x|y} =: \bar{\Sigma}_{x|s},
\end{eqnarray*}
and
\begin{eqnarray}
R(\bm{D}) 
&=& \frac{1}{2} \log\frac{\Sigma_{x|s}}{(\bar{D}^{-1} - \Sigma_{\bar{N}}^{-1} + \Sigma_N^{-1})^{-1}} 
 + \frac{1}{2} \log \frac{\bar{\Sigma}_{x|s}}{\bar{D}} \nonumber \\
 && + \log \frac{(\bar{D}^{-1} - \Sigma_{\bar{N}}^{-1} + \Sigma_N^{-1})^{-1}}{(B^* + \Sigma_N^{-1})^{-1}},
 \label{eq:r-d-for-symmetric}
\end{eqnarray}
where 
\begin{eqnarray*}
B^* = \max[ \bar{D}^{-1} - \Sigma_{\bar{N}}^{-1}, D^{-1} - \Sigma_N^{-1} ].
\end{eqnarray*}
We also have
\begin{eqnarray}
\lefteqn{ R^*_1(D,\bar{D}) + R^*_2(\bar{D},D) } \nonumber \\
&=& \log \frac{\bar{\Sigma}_{x|s}}{\bar{D}} 
 + \log \frac{(\bar{D}^{-1} - \Sigma_{\bar{N}}^{-1} + \Sigma_N^{-1})^{-1}}{(B^* + \Sigma_N^{-1})^{-1}}.
 \label{eq:component-wise-symmetric}
\end{eqnarray}
The lower bound in (\ref{eq:loose-enhanced-bound}) is given by
\begin{eqnarray}
&& \frac{1}{2} \log \frac{\bar{\Sigma}_{x|s}}{\bar{D}} 
+ \frac{1}{2} \log \frac{(\bar{D}^{-1} - \Sigma_{\bar{N}}^{-1} + \Sigma_N^{-1})^{-1}}{(B^* + \Sigma_N^{-1})^{-1}} \nonumber \\
&& ~~~~~~~~~~~~~+ \frac{1}{2} \log \frac{\Sigma_{x|s}}{D_{\min}},
 \label{eq:lower-symmetric-case}
\end{eqnarray}
where $D_{\min} := \min\{ D, \bar{D} \}$.

\begin{figure}[t]
\centering
\includegraphics[width=\linewidth]{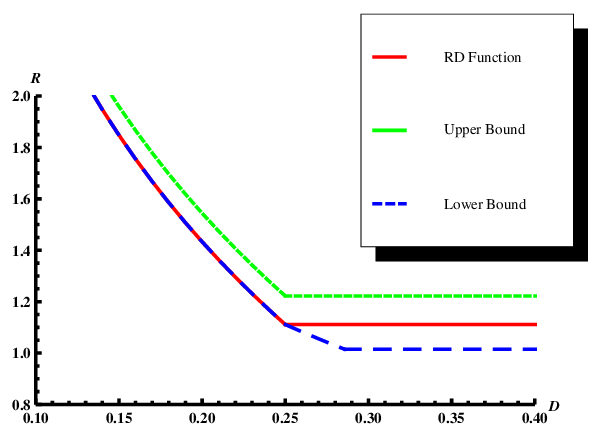}
\caption{
The red solid curve is the rate-distortion function.
The green dotted curve is the upper bound derived by the component-wise scheme.
The blue dashed curve is the lower bound derived by the straightforward application of the enhancement. 
}
\label{Fig:Gaussian-example}
\end{figure}

The distortion such that (\ref{eq:gaussian-matching-condition-enhancement-1}) 
and (\ref{eq:gaussian-matching-condition-enhancement-2}) hold with equality is given by
\begin{eqnarray*}
D^* := (\bar{D}^{-1} - \Sigma_{\bar{N}}^{-1} + \Sigma_N^{-1})^{-1}.
\end{eqnarray*}
For fixed $\bar{D}$, the rate-distortion function, the upper bound, and the 
lower bound are functions of $D$.
From (\ref{eq:r-d-for-symmetric}) 
and (\ref{eq:component-wise-symmetric}), we can find that 
the rate-distortion function and the upper bound are constant
for $D \ge D^*$.
On the other hand, from (\ref{eq:lower-symmetric-case}), we can find that 
the lower bound is constant for $D \ge \bar{D}$.
For $\Sigma_X = 1$, $\Sigma_N = 1$, $\Sigma_{\bar{N}} = 2$,
and $\bar{D} = \frac{2}{7}$, we plot the rate-distortion function,
the upper bound, and the lower bound
in Fig.~\ref{Fig:Gaussian-example}. In this case, note that $D^* = \frac{1}{4}$.
We can find that the upper bound is loose for every $D$,
and that the lower bound is loose for $D > D^*$.

\section{Conclusion}

In this paper, we investigated 
the lossy coding problem for a
product of two sources with two decoders,
and characterized the rate-distortion function.

It is important to extend our result to the case in which
there exists correlation between component sources.
One of such examples is vector Gaussian sources.
As was mentioned in Remark \ref{remark:enhancement},
the converse proof in this paper is motivated by 
the enhancement argument introduced by 
Weingarten {\em et. al.} \cite{weingarten:06}.
However, as we have exemplified in Section \ref{sec:exmaple},
the bound derived by the straightforward application of the 
enhancement argument is loose in general.
Thus, some ingenious way of enhancement might be needed
to solve the vector Gaussian Heegard and Berger problem.
This topic will be investigated in elsewhere.

\section*{Acknowledgment}

The author would like to thank 
Prof.~Tomohiko Uyematsu and 
Prof.~Hirosuke Yamamoto for valuable comments.
He also would like to thank Prof.~Yasutada Oohama
for valuable discussions. He also appreciate
the comments from anonymous reviewers
of ISIT2011, especially for pointing out the result in
Section \ref{remark:exor-construction}.
The comments from anonymous reviewers 
of this journal are also appreciated, which
substantially improved the clarity of this paper.

\appendix
\subsection{Proof of Proposition \ref{proposition-matched-case}}
\label{appendix:proof-of-prposition-matched}

Since the direct part directly follows from
Proposition \ref{proposition:heegard-direct}, we only
prove the converse part.
We proved the converse  part for $R(\bm{D})$.
Suppose that $R$ is $\bm{D}$-achievable. Then, for any $\gamma > 0$,
there exists a code $(\varphi,\phi,\psi)$ satisfying
(\ref{eq:proof-0})--(\ref{eq:distortion-proof-2}),
where we use the same notation as in Section \ref{subsec:proof-converse}.
We will lower bound $H(S_n)$ by using Lemma \ref{lemma:degraded}.
Let $T_n = S_n$, $(A_1,B_1,C_1) = (X_1,Y_1,Z_1)$ and 
$(A_2,B_2,C_2) = (X_2,Y_2,Z_2)$. Then, from Lemma \ref{lemma:degraded},
we have
\begin{eqnarray*}
\lefteqn{
\frac{1}{n}H(S_n)
} \\
&\ge& \frac{1}{n}\sum_{t=1}^n \left[ 
  I(S_n,Y_{1t}^-,Z_{1t}^-,Z_{1t}^+,Z_2^n; X_{1t}|Z_{1t})
 \right. \\
&& + I(Y_{1t}^+,Y_2^n; 
 X_{1t}|Y_{1t},S_n,Y_{1t}^-,Z_{1t}^-,Z_{1t}^+,Z_2^n) \\
&& + I(S_n,Y_1^n,Y_{2t}^-,Z_1^n,Z_{2t}^-,Z_{2t}^+; X_{2t}|Z_{2t}) \\
&& \left. + I(Y_{2t}^+; 
    X_{2t}|Y_{2t},S_n,Y_1^n,Y_{2t}^-,Z_1^n,Z_{2t}^-,Z_{2t}^+) \right] \\
&=& \frac{1}{n}\sum_{t=1}^n \left[ 
  I(W_{1t}; X_{1t}|Z_{1t}) + I(U_{1t};X_{1t}|Y_{1t},W_{1t}) 
 \right. \\
&& \left. + I(W_{2t};X_{2t}|Z_{2t}) + I(U_{2t};X_{2t}|Y_{2t},W_{2t}) 
  \right] \\
&=& I(W_{1T};X_{1T}|Z_{1T},T) + I(U_{1T};X_{1T}|Y_{1T},W_{1T},T) \\
&& + I(W_{2T};X_{2T}|Z_{2T},T) + I(U_{2T};X_{2T}|Y_{2T},W_{2T},T) \\
&=& I(W_{1T},T; X_{1T}|Z_{1T}) + I(U_{1T};X_{1T}|Y_{1T},W_{1T},T) \\
&& + I(W_{2T},T;X_{2T}|Z_{1T}) + I(U_{2T};X_{2T}|Y_{2T},W_{2T},T), 
\end{eqnarray*}
where we set
\begin{eqnarray*}
W_{1t} &=& (S_n,Y_{1t}^-,Z_{1t}^-,Z_{1t}^+,Z_2^n), \\
U_{1t} &=& (Y_{1t}^+,Y_2^n), \\
W_{2t} &=& (S_n,Y_1^n,Y_{2t}^-,Z_1^n,Z_{2t}^-,Z_{2t}^+), \\
U_{2t} &=& Y_{2t}^+,
\end{eqnarray*}
and $T$ is the uniform random number on $\{1,\ldots,n\}$
that are independent of the other random variables.
Note that $W_{1t}, U_{1t}, W_{2t}, U_{2t}$ satisfy
$(W_{it},U_{it}) \markov X_{it} \markov (Y_{it},Z_{it})$
for $i=1,2$.

In a similar reason as in Section \ref{subsec:proof-converse},
there exist functions 
$\hat{X}_i(W_{iT},T,U_{iT},Y_{iT})$ and
$\tilde{X}_i(W_{iT},T,Z_{iT})$ satisfying
(\ref{eq:distortion-condition-1}) 
and (\ref{eq:distortion-condition-2})
for $i=1,2$.
Thus, by taking $W_1=(W_{1T},T)$,
$U_1=U_{1T}$, $W_2 = (W_{2T},T)$,
and $U_2=U_{2T}$, we have that 
there exist $W_1,W_2,U_1,U_2$ satisfying
(\ref{eq:distortion-condition-1}) 
and (\ref{eq:distortion-condition-2}) and 
\begin{eqnarray*}
R(\bm{D})
 &\ge& I(W_1;X_1|Z_1) + I(U_1;X_1|Y_1,W_1) \\
 && + I(W_2;X_2|Z_2) + I(U_2;X_2|Y_2,W_2).
\end{eqnarray*}
Although the auxirially random variables 
$(W_1,U_1,X_1,Y_1,Z_1)$ and $(W_2,U_2,X_2,Y_2,Z_2)$ chosen above are
not necessarily independent of each other, they never 
appear in any term simultaneously. Thus we can take
$(W_1,U_1,X_1,Y_1,Z_1)$ and $(W_2,U_2,X_2,Y_2,Z_2)$ to be independent of each other. 
By using the support lemma \cite{csiszar-korner:81},
we have the statement on the cardinalities of the auxiliary alphabets.
Since $\gamma > 0$ is arbitrary,
by the continuity of $R^*(\bm{D})$ with respect
to $\bm{D}$, we have the converse part for $R(\bm{D})$.
The converse part for $R_{\rom{sum}}(\hat{D},\tilde{D})$ can 
be proved almost in the same manner.
\qed

\subsection{Proof of Lemma \ref{lemma:degraded}}
\label{appendix:proof-of-lemma-degraded}

The lemma is proved in a similar manner as 
Heegard and Berger's converse argument. Our strategy is to
regard $(A^n,B^n,C^n)$ as correlated sources of block length $2n$.
Then, we use Heegard and Berger's converse argument to the
independently but not identical distributed sources of length $2n$. 

First, by chain rules, we have
\begin{eqnarray*}
\lefteqn{
H(T_n)
} \\
&\ge&
I(T_n; A^n|C^n) \\
&=& I(T_n,B^n; A^n|C^n) - I(B^n; A^n|T_n,C^n) \\
&=&
 \sum_{t=1}^n \left[
 I(T_n,B_1^n,B_2^n; A_{1t}|A_{1t}^-,C_1^n,C_2^n) 
\right. \\
&& - I(B_{1t}; A_1^n,A_2^n|T_n,B_{1t}^-,C_1^n,C_2^n) \\
&& + I(T_n,B_1^n,B_2^n; A_{2t}| A_1^n,A_{2t}^-,C_1^n,C_2^n) \\
&& \left. - I(B_{2t}; A_1^n,A_2^n| T_n,B_1^n,B_{2t}^-,C_1^n,C_2^n)
 \right].
\end{eqnarray*}
Since $(A_{1t},C_{1t})$ and $(A_{1t}^-,C_{1t}^-,C_{1t}^+,C_2^n)$
are independent, we have
\begin{eqnarray*}
\lefteqn{
I(T_n,B_1^n,B_2^n; A_{1t}| A_{1t}^-,C_1^n,C_2^n) 
} \\
&=& I(T_n,A_{1t}^-,B_1^n,B_2^n,C_{1t}^-,C_{1t}^+,C_2^n; A_{1t}|C_{1t}) \\
&\ge& I(T_n,B_1^n,B_2^n,C_{1t}^-,C_{1t}^+,C_2^n; A_{1t}|C_{1t}).
\end{eqnarray*}
Similarly, since $(A_{2t},C_{2t})$ and $(A_1^n,A_{2t}^-,C_1^n,C_{2t}^-,C_{2t}^+)$
are independent, we have
\begin{eqnarray*}
\lefteqn{
I(T_n,B_1^n,B_2^n;A_{2t}| A_1^n,A_{2t}^-,C_1^n,C_2^n) 
} \\
&\ge&
 I(T_n,B_1^n,B_2^n,C_1^n,C_{2t}^-,C_{2t}^+; A_{2t}| C_{2t}).
\end{eqnarray*}
Furthermore, since the Markov chains
\begin{eqnarray*}
B_{1t} \markov (A_{1t},C_{1t}) \markov
 (T_n,A_{1t}^-,A_{1t}^+,A_2^n,B_{1t}^-,C_{1t}^-,C_{1t}^+,C_2^n)
\end{eqnarray*}
and
\begin{eqnarray*}
B_{2t} \markov (A_{2t},C_{2t}) \markov
 (T_n,A_1^n,A_{2t}^-,A_{2t}^+,B_1^n,B_{2t}^-,C_1^n,C_{2t}^-,C_{2t}^+)
\end{eqnarray*}
hold, we have
\begin{eqnarray*}
\lefteqn{
I(B_{1t}; A_1^n,A_2^n| T_n,B_{1t}^-,C_1^n,C_2^n)
} \\
&=& I(B_{1t}; A_{1t}| T_n,B_{1t}^-,C_1^n,C_2^n)
\end{eqnarray*}
and 
\begin{eqnarray*}
\lefteqn{
I(B_{2t}; A_1^n,A_2^n| T_n,B_1^n,B_{2t}^-,C_1^n,C_2^n)
} \\
&=& I(B_{2t}; A_{2t}| T_n,B_1^n,B_{2t}^-, C_1^n,C_2^n).
\end{eqnarray*}
Thus, we have
\begin{eqnarray}
\lefteqn{
H(T_n)
} \nonumber \\
&\ge&
 \sum_{t=1}^n \left[
 I(T_n,B_1^n,B_2^n,C_{1t}^-,C_{1t}^+,C_2^n; A_{1t}|C_{1t})
 \right. \nonumber \\
&& - I(B_{1t}; A_{1t}|T_n,B_{1t}^-,C_1^n,C_2^n) \nonumber \\
&& + I(T_n,B_1^n,B_2^n,C_1^n,C_{2t}^-,C_{2t}^+; A_{2t}|C_{2t}) \nonumber 
 \\
&& \left. - I(B_{2t}; A_{2t}| T_n,B_1^n, B_{2t}^-, C_1^n,C_2^n)
 \right].
\label{eq:lemma-proof-1}
\end{eqnarray}
By chain rules, we have
\begin{eqnarray}
\lefteqn{
I(T_n,B_1^n,B_2^n,C_{1t}^-,C_{1t}^+,C_2^n; A_{1t}| C_{1t})
} \nonumber \\
&& - I(B_{1t}; A_{1t}| T_n,B_{1t}^-,C_1^n,C_2^n) \nonumber \\
&=& I(T_n,B_{1t}^-,C_{1t}^-,C_{1t}^+,C_2^n; A_{1t}| C_{1t}) \nonumber \\
&& + I(B_{1t}; A_{1t}| T_n,B_{1t}^-,C_1^n,C_2^n) \nonumber \\
&& + I(B_{1t}^+,B_2^n; A_{1t}| B_{1t},T_n,B_{1t}^-,C_1^n,C_2^n) 
 \nonumber \\
&& - I(B_{1t}; A_{1t}| T_n,B_{1t}^-,C_1^n,C_2^n) \nonumber \\
&=& I(T_n,B_{1t}^-,C_{1t}^-,C_{1t}^+,C_2^n; A_{1t}| C_{1t}) \nonumber \\
&& + I(B_{1t}^+,B_2^n; A_{1t}| B_{1t},T_n,B_{1t}^-,C_1^n,C_2^n).
\label{eq:lemma-proof-2}
\end{eqnarray}
Similarly, we have
\begin{eqnarray}
\lefteqn{
I(T_n,B_1^n,B_2^n,C_1^n,C_{2t}^-,C_{2t}^+; A_{2t}| C_{2t})
} \nonumber \\
&& - I(B_{2t}; A_{2t}| T_n,B_1^n,B_{2t}^-, C_1^n,C_2^n) \nonumber \\
&=& I(T_n,B_1^n,B_{2t}^-,C_1^n,C_{2t}^-,C_{2t}^+; A_{2t}|C_{2t}) 
 \nonumber \\
&& + I(B_{2t}^+; A_{2t}| B_{2t}, T_n, B_1^n,B_{2t}^-,C_1^n,C_2^n).
\label{eq:lemma-proof-3}
\end{eqnarray}
From (\ref{eq:matched-degraded-redescription}), we have
\begin{eqnarray*}
C_{1t} \markov B_{1t} \markov
 (T_n,A_{1t},B_{1t}^-,C_{1t}^-,C_{1t}^+,C_2^n).
\end{eqnarray*}
Thus, we have
\begin{eqnarray}
\lefteqn{
I(B_{1t}^+; A_{1t}| B_{1t},T_n,B_{1t}^-,C_1^n,C_2^n)
} \nonumber \\
&=& I(B_{1t}^+,C_{1t}; A_{1t}| B_{1t},T_n, 
 B_{1t}^-,C_{1t}^-,C_{1t}^+,C_2^n) \nonumber \\
&\ge& I(B_{1t}^+;A_{1t}| B_{1t},T_n,B_{1t}^-,C_{1t}^-,C_{1t}^+,C_2^n).
\label{eq:lemma-proof-4}
\end{eqnarray}
Similarly, from (\ref{eq:matched-degraded-redescription}), we have
\begin{eqnarray}
\lefteqn{
I(B_{2t}^+; A_{2t}| B_{2t}, T_n,B_1^n,B_{2t}^-, C_1^n,C_2^n)
} \nonumber \\
&\ge& I(B_{2t}^+; A_{2t}| B_{2t},T_n,B_1^n,B_{2t}^-,C_1^n,C_{2t}^-,C_{2t}^+).
\label{eq:lemma-proof-5}
\end{eqnarray}
Finally, by substituting (\ref{eq:lemma-proof-2})--(\ref{eq:lemma-proof-5})
into (\ref{eq:lemma-proof-1}), we have
\begin{eqnarray*}
H(T_n)
&\ge&
\sum_{t=1}^n \left[ 
 I(T_n, B_{1t}^-, C_{1t}^-,C_{1t}^+,C_2^n ; A_{1t}|C_{1t}) \right. \\
 && + I(B_{1t}^+, B_2^n ; A_{1t} | B_{1t}, T_n, B_{1t}^-,C_{1t}^-,C_{1t}^+, 
 C_2^n) \\
 && + I(T_n, B_1^n, B_{2t}^-, C_1^n, C_{2t}^-, C_{2t}^+ ; A_{2t} | C_{2t}) \\
 && \left. 
 + I(B_{2t}^+ ; A_{2t} | B_{2t},  T_n, B_1^n, B_{2t}^-, C_1^n, C_{2t}^-, 
  C_{2t}^+ ) \right].
\end{eqnarray*}
\qed

\subsection{Proof of Cardinality Bounds}
\label{appendix:cardinality-bounds}

We prove the cardinality bounds 
by using the support lemma \cite{csiszar-korner:11, elgamal-kim-book}.
We prove by two steps. In the first step, we reduce the cardinality of $W_1$ and $W_2$.
We consider $|{\cal X}_1| + 3$
continuos functions of $P_{U_1X_1|W_1}(\cdot,\cdot|w_1)$ as follows:
\begin{eqnarray*}
f_{0,x_1}(P_{U_1X_1|W_1}(\cdot,\cdot| w_1)) 
	&:=& \sum_{u_1} P_{U_1X_1|W_1}(u_1,x_1|w_1), 
\end{eqnarray*}
for $x_1 = 1,\ldots, |{\cal X}_1| - 1$ and
\begin{eqnarray*}
f_{I,Y}(P_{U_1X_1|W_1}(\cdot,\cdot| w_1)) 
	&:=& H(X_1|Y_1) \\
	&& \hspace{-35mm} - H(X_1|Y_1,W_1=w_1)  + I(U_1;X_1|Y_1, W_1 = w_1), \\
f_{I,Z}(P_{U_1X_1|W_1}(\cdot,\cdot| w_1)) 
	&:=& H(X_1|Z_1) \\
	&& \hspace{-35mm} - H(X_1|Z_1,W_1=w_1) + I(U_1; X_1|Y_1,W_1=w_1), \\
f_{\hat{d}}(P_{U_1X_1|W_1}(\cdot,\cdot| w_1)) 
	&:=& \mathbb{E}\left[ \hat{d}_1(X_1, \hat{X}_1(w_1,U_1,Y_1)) \right], \\
f_{\tilde{d}}(P_{U_1X_1|W_1}(\cdot,\cdot| w_1)) 
	&:=& \mathbb{E}\left[ \tilde{d}(X_1, \tilde{X}_1(w_1,Z_1)) \right].
\end{eqnarray*}
By using the support lemma to these functions, there exists random variable
$W_1^\prime$ with cardinality $|{\cal W}_1^\prime| \le |{\cal X}_1| + 3$ and
the corresponding random variable $U_1^\prime$,
i.e., $P_{U_1^\prime W_1^\prime X_1}(u_1,w_1,x_1) = P_{W_1^\prime}(w_1) P_{U_1X_1|W_1}(u_1,x_1|w_1)$, such that 
the marginal $P_{X_1}$ is preserved and
\begin{eqnarray*}
\lefteqn{ I(W_1^\prime;X_1 | Y_1) + I(U_1^\prime; X_1|Y_1,W_1^\prime) } \\
&=& I(W_1; X_1|Y_1) + I(U_1; X_1|Y_1, W_1) \\
\lefteqn{ I(W_1^\prime; X_1|Z_1) + I(U_1^\prime; X_1|Y_1, W_1^\prime) } \\
&=& I(W_1; X_1|Z_1) + I(U_1; X_1| Y_1, W_1), \\
\lefteqn{ \mathbb{E}\left[ \hat{d}_1(X_1,\hat{X}_1(W_1^\prime,U_1^\prime,Y_1)) \right] } \\
&=& \mathbb{E}\left[ \hat{d}_1(X_1,\hat{X}_1(W_1,U_1,Y_1)) \right], \\
\lefteqn{ \mathbb{E}\left[ \tilde{d}_1(X_1,\tilde{X}_1(W_1^\prime,Z_1)) \right] } \\
&=& \mathbb{E}\left[ \tilde{d}_1(X_1, \tilde{X}_1(W_1,Z_1)) \right].
\end{eqnarray*}
Similarly, there exists $W_2^\prime$ with cardinality $|{\cal W}_2^\prime| \le |{\cal X}_2| + 3$
and the corresponding random variable $U_2^\prime$ such that 
$P_{X_2}$ is preserved and
\begin{eqnarray*}
\lefteqn{ I(W_2^\prime;X_2 | Y_2) + I(U_2^\prime; X_2|Z_2,W_2^\prime) } \\
&=& I(W_2; X_2|Y_2) + I(U_2; X_2|Z_2, W_2) \\
\lefteqn{ I(W_2^\prime; X_2|Z_2) + I(U_2^\prime; X_2|Z_2, W_2^\prime) } \\
&=& I(W_2; X_2|Z_2) + I(U_2; X_2| Z_2, W_2), \\
\lefteqn{ \mathbb{E}\left[ \hat{d}_2(X_2,\hat{X}_2(W_2^\prime,Y_2)) \right] } \\
&=& \mathbb{E}\left[ \hat{d}_2(X_2,\hat{X}_2(W_2,Y_2)) \right], \\
\lefteqn{ \mathbb{E}\left[ \tilde{d}_2(X_2,\tilde{X}_2(W_2^\prime,U_2^\prime,Z_2)) \right] } \\
&=& \mathbb{E}\left[ \tilde{d}_2(X_2, \tilde{X}_2(W_2,U_2,Z_2)) \right].
\end{eqnarray*}

In the next step, we reduce the cardinality of $U_1^\prime$ and $U_2^\prime$.
We consider $|{\cal W}_1^\prime| \cdot |{\cal X}_1| + 1$ 
continuous functions of $P_{W_1^\prime X_1|U_1^\prime}(\cdot,\cdot|u_1)$ as follows: 
\begin{eqnarray*}
g_{0,w_1,x_1}(P_{W_1^\prime X_1|U_1^\prime}(\cdot,\cdot|u_1)) = P_{W_1^\prime X_1|U_1^\prime}(w_1,x_1|u_1)
\end{eqnarray*}
for $(w_1,x_1) = 1,\ldots,|{\cal W}_1^\prime| \cdot |{\cal X}_1| -1$ and
\begin{eqnarray*}
\lefteqn{ g_I(P_{W_1^\prime X_1|U_1^\prime}(\cdot,\cdot|u_1)) } \\
&=& H(X_1|Y_1,W_1^\prime) - H(X_1|Y_1,W_1^\prime,U_1 = u_1), \\
\lefteqn{ g_{\hat{d}}(P_{W_1^\prime X_1|U_1^\prime}(\cdot,\cdot|u_1)) } \\
&=& \mathbb{E}\left[ \hat{d}_1(X_1,\hat{X}_1(W_1^\prime,u_1,Y_1)) \right].
\end{eqnarray*}
By using the support lemma to these functions, there exists $U_1^{\prime\prime}$ with
cardinality $|{\cal U}_1^{\prime\prime}| \le |{\cal W}_1^\prime| \cdot |{\cal X}_1| + 1 = |{\cal X}_1|(|{\cal X}_1| + 3) + 1$ such that
the marginal $P_{W_1^\prime X_1}$ is preserved,
\begin{eqnarray*}
I(U_1^{\prime\prime} ; X_1 | Y_1, W_1^\prime) = I(U_1^\prime; X_1 | Y_1, W_1^\prime),
\end{eqnarray*}
and
\begin{eqnarray*}
\lefteqn{ \mathbb{E}\left[ \hat{d}_1(X_1,\hat{X}_1(W_1^\prime,U_1^{\prime\prime},Y_1)) \right] } \\
&=& \mathbb{E}\left[ \hat{d}_1(X_1,\hat{X}_1(W_1^\prime,U_1^\prime,Y_1)) \right].
\end{eqnarray*}
Similarly, there exists $U_2^{\prime\prime}$ with cardinality $|{\cal U}_2^{\prime\prime}| \le |{\cal X}_2|(|{\cal X}_2| + 3) + 1$
such that $P_{W_2^\prime X_2}$ is preserved, 
\begin{eqnarray*}
I(U_2^{\prime\prime} ; X_2|Z_2, W_2^\prime) = I(U_2^\prime ; X_2 | Z_2, W_2^\prime), 
\end{eqnarray*}
and
\begin{eqnarray*}
\lefteqn{ \mathbb{E}\left[ \tilde{d}_2(X_2,\tilde{X}_2(W_2^\prime,U_2^{\prime\prime},Z_2)) \right] } \\
&=& \mathbb{E}\left[ \tilde{d}_2(X_2, \tilde{X}_2(W_2^\prime,U_2^\prime,Z_2)) \right].
\end{eqnarray*}
By relabeling $(W_1^\prime,U_1^{\prime\prime},W_2^\prime,U_2^{\prime\prime})$
as $(W_1,U_1,W_2,U_2)$, we have the cardinality bounds. \qed

\subsection{Proof of (\ref{eq:end-point-1}) and (\ref{eq:end-point-2})}
\label{appendix-proof-end-point}

\paragraph{Proof of (\ref{eq:end-point-1})}

By noting that $G_p(\cdot)$ is a non-negative and convex function, 
for any $(\check{D},\alpha,\beta,\theta,\tau) \in {\cal Q}_p(\hat{D},\tilde{D})$ we have
\begin{eqnarray*}
B_p(\check{D},\alpha, \beta, \theta,\tau) &=& (\theta - \tau) G_p(\alpha) + \tau G_p(\beta) \\
&& + (1-\theta) G_p(\gamma(\check{D},\alpha,\beta,\theta,\tau)) \\
&\ge& (\theta - \tau) G_p(\alpha) + \tau G_p(\beta) \\
&\ge& \theta G_p\left( \frac{\theta - \tau}{\theta} \alpha + \frac{\tau}{\theta} \beta \right) \\
&\ge& R_p^{WZ}(\hat{D}).
\end{eqnarray*}
Thus, the lefthand side of (\ref{eq:end-point-1}) is larger than or equal to
the righthand side.
On the other hand, for $\tilde{D} = \frac{1}{2}$, by setting 
$\check{D} = \tilde{D}$, $\alpha = \beta$ and
$\tau = \frac{\theta}{2}$, and optimizing $(\beta,\theta)$, 
we can show that the lefthand  side achieves the righthand side
in (\ref{eq:end-point-1}).

\paragraph{Proof of (\ref{eq:end-point-2})}

By noting that $G_p(\cdot)$ is a convex function,
for any $(\check{D}, \alpha,\beta,\theta,\tau) \in {\cal Q}_p(\hat{D},\tilde{D})$ we have
\begin{eqnarray*}
B_p(\check{D},\alpha, \beta, \theta,\tau) &=& (\theta - \tau) G_p(\alpha) + \tau G_p(\beta) \\
&& + (1-\theta) G_p(\gamma(\check{D},\alpha,\beta,\theta,\tau)) \\
&=& (\theta - \tau) G_p(1- \alpha) + \tau G_p(\beta) \\
&& + (1-\theta) G_p(\gamma(\check{D},\alpha,\beta,\theta,\tau)) \\
&\ge& G_p(\check{D}).
\end{eqnarray*}
Since $1 - h(p * \check{D})$ and $G_p(\check{D})$ are monotone
decreasing for $\check{D} \le \frac{1}{2}$, the lefthand side of (\ref{eq:end-point-2})
is larger than or equal to the righthand side. 
On the other hand,
when $\hat{D} = p$, by setting $\check{D} = \tilde{D}$, $\theta = \tau = \alpha = \beta = 0$, we can show that
the lefthand side coincides with the righthand side in (\ref{eq:end-point-2}).
When $\hat{D} = \tilde{D}$, by setting $\check{D} = \tilde{D}$, $\theta = \tau = 1$, $\alpha = 0$, and $\beta = \tilde{D}$,
we can show that the lefthand side coincides with the righthand side in (\ref{eq:end-point-2}).

\subsection{Proof of Theorem \ref{theorem:binary-hamming}}
\label{appendix:proof-of-theorem:binary-hamming}

First, note that (\ref{eq:another-form-of-hb}) can be written as
\begin{eqnarray*}
1 - H(Y|W) + H(Y|U,W) - H(X|U,W),
\end{eqnarray*}
where we used the relations
\begin{eqnarray}
\label{eq:fundamental-relations-1}
I(W;Y) &=& 1 - H(Y|W), \\
I(U,W; X|Y) &=& H(Y|U,W) - H(X|U,W).
\label{eq:fundamental-relations-2}
\end{eqnarray}
To prove (\ref{eq:tian-diggavi}), Tian and Diggavi essentially showed the following
in \cite[Appendix 5]{tian:07}.
\begin{lemma}
\label{lemma:tian-diggavi-bounding}
Let $(U,W)$ be auxiliary random variables satisfying the conditions \ref{condition-single-binary-1} and \ref{condition-single-binary-2}
right after (\ref{eq:rd-may-be-absent}). 
Then, we have
\begin{eqnarray*}
1 - H(Y|W) &\ge& 1 - h(\check{D} *p), \\
H(Y|U,W) - H(X|U,W)
&\ge& B_p(\check{D},\alpha,\beta,\theta,\tau)
\end{eqnarray*}
for some $(\check{D}, \alpha,\beta,\theta,\tau) \in {\cal Q}_p(\hat{D},\tilde{D})$.
\end{lemma} 

By noting that $Z_1$ and $Y_2$ are constant and by using chain rules, 
for a fixed auxiliary random variable $(U_1,W_1,U_2,W_2)$,
we can rewrite the rate condition of Theorem \ref{theorem-main} as
\begin{eqnarray*}
\lefteqn{ \max\{ I(U_1, W_1; X_1|Y_1) + I(W_2;Z_2) + I(U_2,W_2; X_2|Z_2), } \\
&& I(W_1; Y_1) + I(U_1,W_1; X_1|Y_1) + I(U_2,W_2; X_2|Z_2) \} .
\end{eqnarray*}
Then, by using Lemma \ref{lemma:tian-diggavi-bounding}
and the relations in (\ref{eq:fundamental-relations-1}) and (\ref{eq:fundamental-relations-2}), we have
\begin{eqnarray*}
\lefteqn{ I(U_1, W_1; X_1|Y_1) + I(W_2;Z_2) + I(U_2,W_2; X_2|Z_2) } \\
&\ge& 1 - h(\check{D}_2 * p_2) + B_{p_1}(\check{D}_1,\alpha_1,\beta_1,\theta_1,\tau_1) \\
&& ~~~~ + B_{p_2}(\check{D}_2,\alpha_2,\beta_2,\theta_2,\tau2)
\end{eqnarray*}
and
\begin{eqnarray*}
\lefteqn{ I(W_1; Y_1) + I(U_1,W_1; X_1|Y_1) + I(U_2,W_2; X_2|Z_2) } \\
&\ge& 1 - h(\check{D}_1 * p_1) + B_{p_1}(\check{D}_1,\alpha_1,\beta_1,\theta_1,\tau_1) \\
&& ~~~~ + B_{p_2}(\check{D}_2,\alpha_2,\beta_2,\theta_2,\tau2)
\end{eqnarray*}
for some
\begin{eqnarray*}
(\check{D}_1,\alpha_1,\beta_1,\theta_1,\tau_1) \in {\cal Q}_{p_1}(\hat{D}_1,\tilde{D}_1)
\end{eqnarray*}
and 
\begin{eqnarray*}
(\check{D}_2, \alpha_2,\beta_2,\theta_2,\tau_2) \in {\cal Q}_{p_2}(\tilde{D}_2,\hat{D}_2).
\end{eqnarray*}
Thus, the lefthand side is larger than or equal to the righthand side
in (\ref{eq:binary-example-expocit-form}). We can prove the other
direction of inequality by using the reverse test channels described in
Figs.~\ref{Fig:test-channel-w12} and \ref{Fig:test-channel-wx}.

\bibliographystyle{../09-04-17-bibtex/IEEEtran}
\bibliography{../09-04-17-bibtex/reference.bib}


\begin{IEEEbiographynophoto}{Shun Watanabe}
(M'09) received the B.E., 
M.E., and Ph.D.\ degrees from Tokyo Institute of Technology
in 2005, 2007, and 2009 respectively. Since April 2009, he has been
an Assistant Professor in the Department of Information 
Science and Intelligent Systems of  the University of Tokushima.
Since April 2013, he also has been a visiting Assistant Professor
in the Institute for Systems Research of the University of Maryland, College Park.
His current research interests are in the areas of
information theory, quantum information theory,
and quantum cryptography.
\end{IEEEbiographynophoto}

\end{document}